\documentclass[11pt]{article}

\usepackage[margin=1in]{geometry}

\usepackage{mathtools}
\usepackage{hyperref}
\usepackage[noadjust]{cite}
\usepackage[utf8]{inputenc}

\usepackage{xcolor,listings,tcolorbox}

\tcbuselibrary{listings}

\definecolor{cellbg}{HTML}{F7F7F7}       
\definecolor{accentblue}{HTML}{465A75}   
\definecolor{pykeyword}{HTML}{008000}    
\definecolor{pycomment}{HTML}{408080}    
\definecolor{pystring}{HTML}{BA2121}     

\newtcblisting{codebox}[1][]{
    listing only,
    listing options={
        basicstyle=\ttfamily\small,
        columns=fullflexible,
        keepspaces=true,
        breaklines=true,
        showstringspaces=false,
        tabsize=4
    },
    colback=cellbg,
    colframe=accentblue,
    boxrule=0.3pt,
    leftrule=2pt,
    arc=2pt,
    outer arc=2pt,
    top=0pt,
    bottom=0pt,
    left=8pt,
    right=8pt,
    #1
}

\newcommand{\rt}{{\tilde r}}
\newcommand{\meanrt}{\langle \rt \rangle}
\newcommand{\eff}{{\mathrm{eff}}}

\numberwithin{figure}{section}
\numberwithin{equation}{section}

\begin{document}
\pagenumbering{roman}

\begin{titlepage}
\title{\textbf{Exploring continuous \(\beta\)-ensembles:\\A Python implementation for random matrix spectral statistics}\vspace{10mm}}

\author{
\textbf{Dorin Weissman} \\ \href{mailto:dorin.weissman@na.infn.it}{dorin.weissman@na.infn.it}
}

\date{
\emph{INFN Sezione di Napoli,}\\
	\emph{Monte S. Angelo, Via Cintia, 80126 Naples, Italy} 
\\[1.5\baselineskip] \today}
	
\maketitle

\begin{abstract}
We present an open-source Python package for sampling the Gaussian, Laguerre, Jacobi, and Circular $\beta$-ensembles of random matrix theory. The package implements the Dumitriu--Edelman and Killip--Nenciu constructions, allowing efficient generation of random spectra for general $\beta > 0$. In addition to spectrum generation, it includes tools for the analysis of spectral statistics, from standard nearest-neighbor spacings and spacing ratios to non-adjacent $k$-spacings and the spectral form factor. These tools can be applied to generic spectral data, allowing users to compare them with and fit them to $\beta$-ensemble predictions. In this note, we review the $\beta$-ensembles, describe the package interface, and illustrate its use through several numerical experiments motivated by applications to quantum chaos. Our numerical results include an analysis of $\beta$ as a continuous fitting parameter in spacing ratio statistics, an examination of the numerical evidence for the conjectured $k$-spacing ratio distributions, and a study of the spectral form factor for general values of $\beta$.
\end{abstract}

\end{titlepage}

\flushbottom

\tableofcontents

\clearpage
\pagenumbering{arabic}

\section{Introduction}
Random matrix theory is a primary tool in studying quantum chaotic systems. The classical Gaussian ensembles are the Gaussian orthogonal, unitary, and symplectic ensembles (GOE, GUE, and GSE), which are ensembles of random matrices associated with the three canonical symmetry classes. A natural generalization is provided by the Gaussian $\beta$-ensemble, which is a continuous one-parameter family of distributions with the GOE, GUE, and GSE residing at the special points $\beta = 1,2$, and 4.

Similar generalizations also exist for other families of RMT ensembles: the Circular, Laguerre, and Jacobi ensembles. The continuous $\beta$-ensembles are extensively studied in the mathematical literature (see \cite{Forrester:book} for an introduction) but are not widely used in physical applications, as physical systems usually have symmetry constraints which lead them to one of the classical ensembles. On the other hand, continuous $\beta$ is a useful parameter to model systems that are not quite GOE or GUE. At the very least this parameter can quantify how far one is from one of the classical ``fixed points'', and allow for smooth interpolations or transitions between them. In addition, in the limit $\beta\to0$, the level statistics approach the non-chaotic Poisson statistics, while in the $\beta\to\infty$ limit spectra ``crystallize'' and become regular.

The $\beta$-ensembles also have a physical interpretation in terms of Dyson's log-gas model \cite{Dyson1962}. In this picture, the eigenvalues behave as one-dimensional particles with pairwise logarithmic interactions, while $\beta$, often called the Dyson index, has the natural interpretation of an inverse temperature.

Our interest in the $\beta$-ensembles stems from our study of chaos in string theory \cite{Bianchi:2023uby}. In this and follow-up works we showed that data obtained from the scattering amplitudes of highly-excited strings can be well described by theoretical predictions of RMT, but often fitted with non-integer $\beta$. One viewpoint is to consider this a bug: if we could only perfectly clean up our computation to isolate the universal RMT statistics, or go to the infinite $N$ limit of the highly-excited string, we would surely find $\beta = 1$ or 2. The other viewpoint is that this is a feature: in the regime accessible to our calculation, the scattering process may exhibit an effective value of $\beta$ corresponding to neither the orthogonal nor unitary ensemble, but lying somewhere in between. It is quite possible we will get insight into the physics in this regime by studying the interpolating $\beta$-ensembles.

We present an open-source Python package \texttt{beta\_ensembles}, available from \href{https://github.com/dorinw/beta_ensembles}{GitHub} and \href{https://pypi.org/project/beta-ensembles/}{PyPI} and archived on Zenodo \cite{Weissman:2026be}, that provides a set of numerical tools for studying the Gaussian, Laguerre, Jacobi, and Circular $\beta$-ensembles (GBE, LBE, JBE, and CBE) at general $\beta$. There are many widely available tools for working with random matrices, but they often implement exclusively the classical ensembles.\footnote{An exception is DPPy \cite{JMLR:v20:19-179}, which also implements the spectrum-generating algorithms used here, though its primary focus is on determinantal point processes rather than RMT.} Nevertheless, efficient algorithms for generating the spectra of $\beta$-ensembles are widely known. For the Gaussian and Laguerre ensembles we use the Dumitriu--Edelman construction \cite{Dumitriu:2002}, while for the Circular and Jacobi ensembles we use the Killip--Nenciu \cite{Killip:2004} construction. The Gaussian, Laguerre, and Jacobi ensembles admit a construction in terms of real tridiagonal symmetric matrices, whose eigenvalues can be found efficiently using specialized algorithms.

We also include a modest suite of tools for spectral analysis. In particular, they can be used to easily compute and examine nearest-neighbor spacings, spacing ratios, non-adjacent spacings and their ratios, and the spectral form factor.

This note is a companion piece to the Python package. It supplements the documentation by establishing notation and mathematical conventions, and by demonstrating several physics-motivated uses of the package. We hope that it will be of use to researchers interested in working with $\beta$-ensembles or in comparing data with the spectral statistics of $\beta$-ensembles.

The note is structured as follows. We begin in section \ref{sec:ensembles} with a short review, defining the four families of $\beta$-ensembles and presenting the algorithms used to generate their spectra. In section \ref{sec:spectral_statistics} we discuss the spectral statistics, introducing the level spacings and spacing ratios and their expected distributions. We also include the non-adjacent $k$-spacings, $k$-spacing ratios, and the spectral form factor. We write explicitly all the relevant predictions at arbitrary $\beta$.

Section \ref{sec:package_overview} provides an outline of the Python package, describing the main functions and their intended usage, including some notes on performance. Section \ref{sec:demo} demonstrates basic usage by reproducing the expected distributions of eigenvalues and spacings for the Gaussian, Laguerre and Circular ensembles.\footnote{Addendum \ref{app:jacobi} shows the same for the Jacobi ensembles.}

Readers who are interested in $\beta$-ensembles may skip ahead and go directly to section \ref{sec:applications}, which is dedicated to showing applications of our code through a series of numerical studies. We present three ``experiments'', each with a different focus:

\begin{itemize}
    \item In section~\ref{sec:fitting_beta} we begin a systematic study of continuous $\beta$ as a fitting parameter in the simplest case of starting with GBE spectra and fitting their spacing ratios to the expected distribution. We establish what uncertainties there are in the fitted values of $\beta$ and see how different goodness of fit measures behave. We show that the systematic deviation of the spacing ratio distribution from the ABGR formula at finite $N$ can be modeled by fitting an effective value $\beta_\eff$, which is always smaller than the input value of $\beta$.

    \item The second experiment in section \ref{sec:k_spacings} is a short study of $k$-spacing ratios. We verify to what extent the effective index $\beta^{(k)} = \frac{k(k+1)}{2}\beta + (k-1)$ can model the resulting distributions in the Gaussian $\beta$-ensembles. Our main finding is that finite $N$ corrections lead to significant discrepancies from the expected formula as $k$ is increased.

    \item In the third experiment of section \ref{sec:sff_exp}, we revisit the spectral form factor of continuous $\beta$-ensembles, extending the study initiated in \cite{Bianchi:2024fsi}. For $1\leq\beta\leq 4$, we provide additional evidence for the interpolation formulas discovered in \cite{Bianchi:2024fsi},
    \begin{align}
        R_2^{OU} =& (\frac2\beta-1)R_2^\text{GOE}+(2-\frac{2}\beta) R_2^\text{GUE} \\
        R_2^{US} =&(\frac4\beta-1)R_2^\text{GUE}+(2-\frac4\beta)R_2^\text{GSE}
    \end{align}
    which, as far as we know, have not been derived elsewhere in the literature. We remark also on the behavior of the SFF for values of $\beta$ larger than 4, and how it interpolates in a continuous fashion to the large $\beta$ limit.
\end{itemize}

All the numerical results in this note are exactly reproducible from the example notebooks available in the code repository. We hope that the \texttt{beta\_ensembles} package and this presentation will encourage others to continue experimenting and gain more insight into the role that continuous $\beta$-ensembles can play in modeling quantum chaotic systems.

There are a few directions in which our work can be extended. We currently implement the distributions of spacings and spacing ratios, together with their non-adjacent generalizations, but not other statistical measures commonly used in the RMT literature. On the experimental side, our present study focuses almost exclusively on the ``interpolating regime'' $1\leq\beta\leq4$, where $\beta$ lies near the values corresponding to the classical ensembles. As our code supports a much broader range of $\beta$, it can be used to investigate the small- and large-$\beta$ limits.

\clearpage
\section{The Gaussian, Laguerre, Jacobi, and Circular \texorpdfstring{$\beta$}{beta}-ensembles} \label{sec:ensembles}
In this section we introduce the Gaussian, Laguerre, Jacobi, and Circular $\beta$-ensembles, or GBE, LBE, JBE, and CBE, respectively. This is not intended as an all-purpose review. For an introduction to the topic see \cite{Forrester:book}. The main purpose here is to establish our conventions and describe the algorithms used to sample from them.

\subsection{Gaussian ensembles}
The Gaussian $\beta$-ensemble (GBE) can be defined starting from the joint probability distribution function of eigenvalues:
\begin{equation} \label{eq:dist_lambda_GBE}
P_\text{GBE}(\lambda_1,\lambda_2,\ldots,\lambda_N) = {\cal N}_\text{GBE}(\beta,N) \times \exp\left(-\frac\beta4\sum_{n=1}^N \lambda_n^2\right) \prod_{1\leq m < n \leq N} |\lambda_m-\lambda_n|^\beta  \end{equation}
where ${\cal N}_\text{GBE}(\beta,N)$ is a known normalization constant.

This defines a distribution for any real $\beta > 0$. The classical Gaussian ensembles: the orthogonal (GOE), unitary (GUE) and symplectic (GSE), are obtained by fixing $\beta = 1$, 2, and 4, respectively.

An excellent model for the generation of GBE spectra is the construction of Dumitriu and Edelman \cite{Dumitriu:2002}, which defines an ensemble of real, tridiagonal symmetric matrices with random entries. The eigenvalues of these matrices follow exactly the distribution of eq.~\eqref{eq:dist_lambda_GBE}, for any $N$ and $\beta$.

It is important to note that, because of the choice of basis, setting $\beta = 1$, 2, or 4 in the following algorithm will \textit{not} be equivalent to using the GOE, GUE, or GSE, except that the distributions of eigenvalues will coincide. In other words, the distribution of eigenvectors in the classical random matrix ensembles cannot be inferred from the tridiagonal construction.

To generate GBE spectra, we proceed with the Dumitriu-Edelman model as follows.\footnote{Note that we use a different normalization convention of the eigenvalues than in the original reference \cite{Dumitriu:2002}.} We define the $N\times N$ matrices of the symmetric tridiagonal form
\begin{equation} \label{eq:M_beta_GBE}
{\cal M} = \frac{1}{\sqrt{\beta}}\left(\begin{matrix}
d_1 & x_1 & & & \\
x_1 & d_2 & x_2 & &  \\
& \ddots & \ddots & \ddots & \\
& & x_{N-2} & d_{N-1} & x_{N-1} \\
& & & x_{N-1} & d_N
\end{matrix}\right) \end{equation}
The $N$ diagonal elements $d_i$ and the $N-1$ elements on the off-diagonals $x_i$ are independent random variables. The diagonal elements are each taken from the normal distribution with zero mean and $\sigma=\sqrt{2}$,
\begin{equation}
    d_i \sim N(0,\sqrt{2})
\end{equation}
while the off-diagonal elements are each drawn from a $\chi$-distribution with varying parameter as
\begin{equation}
    x_k \sim \chi_{(N-k)\beta}
\end{equation}
The eigenvalues of the matrices $M$ defined in this way obey the statistics of \ref{eq:dist_lambda_GBE}.

\subsection{Laguerre ensembles}
The Laguerre $\beta$-ensemble is defined by the joint probability distribution function:
\begin{equation} \label{eq:dist_lambda_LBE}
    P_\text{LBE}(\lambda_1,\ldots,\lambda_N) = {\cal N}_\text{LBE}(\beta,\alpha,N) \prod_{m=1}^N \lambda_m^\alpha e^{-\frac\beta2\lambda_m} \times\prod_{1\leq m< n \leq N}|\lambda_m-\lambda_n|^\beta
\end{equation}
This distribution has, in addition to the Dyson index $\beta > 0$, the additional parameter $\alpha > -1$.

There are other ways to parametrize the ensemble. Classically, a random matrix in the Laguerre--Wishart ensembles can be defined starting from a rectangular $M\times N$ matrix with Gaussian components, which is multiplied by its transpose to produce the $N\times N$ matrix. The construction above reproduces the statistics of the classical ensembles $\beta = 1$, 2, and 4 (LOE, LUE, and LSE) when mapping the continuous parameter $\alpha$ to the integer $M$ as in
\begin{equation} \label{eq:LBE_alpha}
    \alpha + 1 = \frac\beta2(M-N+1) 
\end{equation}
The constraint $\alpha > -1$ is equivalent to $M \geq N$.

To construct LBE spectra, we follow again a construction provided by Dumitriu--Edelman \cite{Dumitriu:2002}, in this case with bidiagonal matrices as the elementary object. To follow the construction in \cite{Dumitriu:2002} we use their parameters $(a,p)$, which map to ours through the relations
\begin{equation}
    a-p=\alpha \,,\qquad p = 1+\frac\beta2(N-1)
\end{equation}
implying
\begin{equation}
    a = \alpha + 1 + \frac\beta2(N-1)
\end{equation}

To generate spectra distributed according to eq.~\eqref{eq:dist_lambda_LBE}, we construct the bidiagonal matrix:
\begin{equation}
B = \left(\begin{matrix}
d_1 &  & & & \\
x_1 & d_2 &  & &  \\
& \ddots & \ddots & & \\
& & x_{N-2} & d_{N-1} & \\
& & & x_{N-1} & d_N
\end{matrix}\right) \end{equation}
where the diagonal and off-diagonal elements are independent random variables drawn from the $\chi$-distribution according to:
\begin{equation} d_k \sim \chi\big(2a - (k-1)\beta\big) \,, \qquad x_k \sim \chi\big(\beta(N-k)\big)
\end{equation}
Finally, the matrix
\begin{equation}
 {\cal M} = \frac1\beta B B^T
\end{equation}
whose components are given by
\begin{equation} \label{eq:M_beta_LBE}
{\cal M} = \frac1\beta\left(\begin{matrix}
d_1^2 & d_1 x_1 & & & \\
d_1 x_1 & d_2^2 + x_1^2 & d_2x_2 & &  \\
& \ddots & \ddots & \ddots & \\
& & d_{N-2}x_{N-2} & d_{N-1}^2+x_{N-2}^2 & d_{N-1}x_{N-1} \\
& & & d_{N-1} x_{N-1} & d_N^2 +x_{N-1}^2
\end{matrix}\right) \end{equation}
is a symmetric tridiagonal matrix whose eigenvalues follow~\eqref{eq:dist_lambda_LBE}.

\subsection{Jacobi ensembles}
The Jacobi $\beta$-ensembles are defined starting from the joint probability distribution function:
\begin{equation} \label{eq:dist_lambda_JBE}
    P_\text{JBE}(\lambda_1,\ldots,\lambda_N) = {\cal N}_\text{JBE}(\beta,a,b,N) \prod_{m=1}^N (1+\lambda_m)^a (1-\lambda_m)^b\times\prod_{1\leq m< n \leq N}|\lambda_m-\lambda_n|^\beta
\end{equation}
This distribution has, in addition to the Dyson index $\beta > 0$, two parameters $a > -1$ and $b>-1$. 

As with the Laguerre ensembles, there are other conventions. In the one we use, eigenvalues are normalized to always be in the range $\lambda\in[-1,1]$. The classical Jacobi ensembles, also known as the MANOVA ensembles, are defined starting from random matrices of dimensions $M_1 \times N$ and $M_2\times N$. The integer parameters $(M_1,M_2)$ are related to $(a,b)$ by:
\begin{equation} \label{eq:parameters_JBE}
    a + 1 = \frac\beta2(M_1-N+1)\,,\qquad b + 1 = \frac\beta2(M_2-N+1)
\end{equation}
from which it follows that we must have $M_1 \geq N$ and $M_2 \geq N$ to get valid parameters $a > -1$ and $b > -1$, respectively. While the classical ensembles allow for only integer (or half-integer for $\beta=1$) $a$ and $b$, they can be made continuous real parameters.

The tridiagonal matrix model for the JBE was introduced by Killip and Nenciu in \cite{Killip:2004}. We note that there are a few differences in convention and notation between this work and theirs, but the algorithm that follows is the same.

We define the $N\times N$ matrices of the symmetric tridiagonal form
\begin{equation} \label{eq:M_beta_JBE}
J = \frac{1}2\left(\begin{matrix}
d_1 & x_1 & & & \\
x_1 & d_2 & x_2 & &  \\
& \ddots & \ddots & \ddots & \\
& & x_{N-2} & d_{N-1} & x_{N-1} \\
& & & x_{N-1} & d_N
\end{matrix}\right) \end{equation}
The $N$ diagonal elements $d_i$ and the $N-1$ elements on the off-diagonals $x_i$ are random variables. These random variables are constructed by drawing from the Beta distribution. In our convention a variable $X\sim B(s,t)$, has the PDF $p(x)\propto x^{s-1} (1-x)^{t-1}$ for $0< x < 1$. Since \cite{Killip:2004} uses a different convention, to match their prescription we first draw the variables $\gamma_n$ using\footnote{The difference in convention of \cite{Killip:2004} also leads us to reverse the roles of $a$ and $b$.}
\begin{equation}
    \gamma_{k} \sim B\left(\frac{2N-k-2}{4}\beta + b+1, \,\frac{2N-k-2}{4}\beta+a+1\right)
\end{equation}
for even $k = 0,2,\ldots, 2N-2$, and
\begin{equation}
    \gamma_{k} \sim B\left(\frac{2N-k-3}{4}\beta + a + b +2, \,\frac{2N-k-1}{4}\beta\right)
\end{equation}
for odd $k = 1,3,\ldots, 2N-3$. Then we can define the coefficients
\begin{equation}\alpha_k = 1-2\gamma_k \end{equation}
for $k = 0,\ldots,2N-2$ as in \cite{Killip:2004}. We also set $\alpha_{-1} = \alpha_{2N-1} = -1$.

The elements of the random tridiagonal matrix are given by
\begin{equation}
    d_{k+1} = (1-\alpha_{2k-1})\alpha_{2k} - (1+\alpha_{2k-1})\alpha_{2k-2}
\end{equation}
for $k = 0,\ldots,N-1$ and 
\begin{equation}
    x_{k+1} = \sqrt{(1-\alpha_{2k-1})(1-\alpha_{2k}^2)(1+\alpha_{2k+1})}
\end{equation}
for $k = 0,\ldots,N-2$. Then the eigenvalues of the matrix defined in eq.~\eqref{eq:M_beta_JBE} are distributed as in the JBE, i.e. according to eq.~\eqref{eq:dist_lambda_JBE}.

\subsection{Circular ensembles}
The Circular ensembles describe random unitary matrices, whose eigenvalues are all on the unit circle, namely $\lambda_n = e^{i\phi_n}$. 

The joint probability distribution of the \emph{eigenphases} $\{\phi_i\}$ in the Circular $\beta$-ensemble is given by:
\begin{equation} \label{eq:dist_lambda_CBE}
P_\text{CBE}(\phi_1,\phi_2,\ldots \phi_N) = {\cal N}_\text{CBE}(\beta,N) \prod_{1\leq m < n \leq N}|e^{i\phi_m}-e^{i\phi_n}|^\beta
\end{equation}
where ${\cal N}_\text{CBE}(\beta,N)$ is a known normalization constant.

The matrix model that we use to generate spectra obeying the distribution \eqref{eq:dist_lambda_CBE} for arbitrary $\beta$ is due to Killip and Nenciu \cite{Killip:2004}. The construction is more involved than in the Gaussian case and relies on the use of Cantero-Moral-Velazquez (CMV) matrices \cite{Cantero:2003}.

The algorithm is as follows. In order to define a random $N\times N$ unitary matrix from the CBE, we first define $N$ independent complex random variables 
\begin{equation} \alpha_k \equiv \sqrt{\rho_k} e^{i\phi_k} \end{equation}
for $k=0,1,\ldots,N-1$, such that
their phases $\phi_k$ are drawn from a uniform distribution,
\begin{equation} \phi_k \sim U(0,2\pi) \end{equation}
while the radial variables $\rho_k$ are drawn from the Beta-distribution\footnote{We note again that $X\sim B(s,t)$ here means that $x$ has the PDF $p(x)\propto x^{s-1} (1-x)^{t-1}$ for $0< x < 1$, while \cite{Killip:2004} uses a different convention.} as
\begin{equation} \label{eq:KN_rho}
\rho_k \sim B\left(1,\frac12(N-k-1)\beta\right) \end{equation}
for $k = 0,1,\ldots,N-2$. In the case of $k=N-1$ one defines $\rho_{N-1} \equiv 1$.

Then, one defines the auxiliary $2\times2$ matrices
\begin{equation} \Xi_{k} \equiv \left(\begin{matrix}
    \alpha_k^* & \sqrt{1-|\alpha_k|^2} \\
    \sqrt{1-|\alpha_k|^2} & -\alpha_k
\end{matrix}\right)\end{equation}
for $k=0,1,\ldots,N-2$, and two $1\times1$ matrices
\begin{equation} \Xi_{-1} \equiv \big(-1\big)\,,\qquad \Xi_{N-1} \equiv \big(\alpha_{N-1}^*\big) \end{equation}
These are combined into two block diagonal matrices
\begin{equation} {\cal M}_1 \equiv \mathrm{diag}\left(\Xi_{-1},\Xi_{1},\Xi_3,\ldots\right)\,, \qquad {\cal M}_2 \equiv \mathrm{diag}\left(\Xi_0,\Xi_2,\Xi_4,\ldots\right) \end{equation}
Finally, the matrix 
\begin{equation} \label{eq:M_beta_CBE}
    {\cal M} = {\cal M}_1 {\cal M}_2
\end{equation}
(alternatively ${\cal M}_2 {\cal M}_1$) is a random $N\times N$ five-diagonal unitary matrix whose eigenvalues follow the desired distribution of eq.~\eqref{eq:dist_lambda_CBE}, for generic values of $\beta>0$.

\clearpage
\section{Spectral statistics} \label{sec:spectral_statistics}
In this section we collect relevant formulas for the spectral statistics of $\beta$-ensembles. Most of the results cited here are canonical and can be found in any RMT reference, for instance the book \cite{Mehta:book}.

\subsection{Spectral density and unfolding}\label{sec:unfolding}
We defined the GBE, LBE, JBE, and CBE in terms of the multivariate joint distribution functions of their eigenvalues, \eqref{eq:dist_lambda_GBE}, \eqref{eq:dist_lambda_LBE}, \eqref{eq:dist_lambda_JBE} and~\eqref{eq:dist_lambda_CBE} respectively.

The ensembles share similar universal statistics in the large $N$ limit, but each of the different ensembles has a different spectral density function $\rho(\lambda)$, defined as the probability distribution that any eigenvalue will be equal to $\lambda$. Data from a given physical system might have an altogether different spectral density function to our ensembles while still following universal RMT statistics.

To observe the universal features we ``unfold'' our spectra, applying a transformation $f(\lambda)$, that will uncover the universal statistics:
\begin{equation}
    z_n = f(\lambda_n)
\end{equation}
The purpose of unfolding is to change to a new variable $z$, in terms of which the density is uniform in the range $[0,N]$, i.e.:
\begin{equation} \label{eq:uniform_density}
    \rho_z(z) = \begin{cases}
        \frac{1}{N} & 0\leq z\leq N\ \\
        0 & \text{else}
    \end{cases}
\end{equation}

A transformation that will do this for general $\rho(\lambda)$ is the one defined using the cumulative distribution function:
\begin{equation}
    I(\lambda) = N\int_{-\infty}^\lambda \rho(\lambda^\prime)d\lambda^\prime
\end{equation}
If we define
\begin{equation}
    z_n = I(\lambda_n)
\end{equation}
it follows immediately that the density $\rho(z)$ is as in~\eqref{eq:uniform_density}.

Note that while we always normalize $\rho(\lambda)$ as a probability distribution function, we defined $I(\lambda)$ with an extra factor of $N$, such that $z\in[0,N]$. Strictly speaking, the CDF is given by $\mathrm{CDF}(\lambda) = I(\lambda)/N$.

\subsubsection{Circular ensembles}
The Circular ensembles are the simplest in terms of their spectral density function, as rotation symmetry ensures that the distribution of the eigenphases $\phi_n$ is uniform in the range $\phi\in[-\pi,\pi]$. The ``unfolding'' simply amounts to shifting and rescaling the eigenphases:
\begin{equation}
    z_n = \frac{N}{2\pi}(\phi_n+\pi)
\end{equation}
such that the unfolded variable $z$ is in the range $[0,N]$.

\subsubsection{Gaussian ensembles}
The spectra of the Gaussian ensembles will obey the Wigner semicircle law. The average eigenvalue density at large $N$ is expected to converge to:
\begin{equation} \rho(\lambda) = \frac{1}{2\pi N}\sqrt{4N-\lambda^2} \label{eq:dist_semicircle}\end{equation}
Note that with our definition of eq.~\eqref{eq:dist_lambda_GBE}, the radius of the semicircle is always $R = 2\sqrt{N}$ independently of $\beta$, unlike some definitions which use different normalizations for GOE, GUE, and GSE.

To unfold the spectrum we can use the CDF, integrating eq.~\eqref{eq:dist_semicircle}:
\begin{equation} I_{\text{s.c.}}(\lambda) = 
    \begin{cases} 0 & \lambda < -2\sqrt{N} \\
            \frac{N}{2} + \frac{N}{\pi}\arctan\left(\frac{\lambda}{\sqrt{4N-\lambda^2}}\right) + \frac{\lambda\sqrt{4N-\lambda^2}}{4\pi} 
            & -2\sqrt{N}\leq \lambda \leq 2\sqrt N 
             \\            N & \lambda > 2\sqrt N \end{cases}
            \label{eq:dist_semicircle_cdf}
\end{equation}
An important caveat here is that, for finite values of $N$, the formula~\eqref{eq:dist_semicircle} is not exact. In particular, there is a finite probability for eigenvalues to lie outside the range $-2\sqrt{N}\leq \lambda\leq2\sqrt{N}$. This fact combined with the singular nature of the distribution at its endpoints means that the unfolding may introduce edge effects.

One practical workaround is to unfold the spectrum using~\eqref{eq:dist_semicircle_cdf}, but keep only eigenvalues from the bulk of the distribution, \textit{e.g.} by dismissing the smallest and largest 10\% of eigenvalues. We will use this method often in our examples.

\subsubsection{Laguerre ensembles}
The Laguerre ensembles do not have a single large matrix limit, $N\to\infty$, as we have to also consider the scaling of the parameter $\alpha$ (or $M$). The relevant parameter is the ratio:
\begin{equation}
    \gamma \equiv \frac{N}{M} =\frac{N}{N-1+\frac2\beta(\alpha+1)}
\end{equation}
which by definition is $0\leq \gamma \leq 1$. In this limit the eigenvalue density function is described by the Marchenko--Pastur law (sometimes called the quarter-circle law):
\begin{equation} \label{eq:dist_MP}
    \rho_\mathrm{MP}(\lambda;\gamma) = \frac{1}{2\pi N }\sqrt{\frac{(x_+-\lambda)(\lambda-x_-)}{\lambda^2}}
\end{equation}
with
\begin{equation}
    x_\pm = \frac{N}{\gamma}(1 \pm\sqrt{\gamma})^2
\end{equation}
This distribution has support only in the range $x_-\leq\lambda\leq x_+$. In the limit $N\to\infty$ with fixed $\alpha=\alpha_0$, and therefore $\gamma=1$, one gets the distribution:
\begin{equation} \label{eq:dist_MP_1}
    \rho_\mathrm{MP}(\lambda;\gamma=1) = \frac{1}{2\pi N}\sqrt{\frac{4N-\lambda}{\lambda}} 
\end{equation}

As for the GBE, we can write an analytic expression for the CDF that can be used for unfolding the spectra of Laguerre ensembles, by integrating \eqref{eq:dist_MP}:
\begin{equation}
    I_\mathrm{MP}(\lambda) =
    \begin{cases} 0 & \lambda < x_- \\
            \frac{N}{2} + \frac{1}{2\pi}\sqrt{(x_+-\lambda)(\lambda-x_-)} + \frac{N(1+\gamma)}{2\pi\gamma}\arcsin\left(\frac{2\lambda-x_+-x_-}{x_+ - x_-}\right) - & \\\quad-\frac{N(1-\gamma)}{2\pi\gamma}\arcsin\left(\frac{(x_++x_-)\lambda-2x_+x_-}{(x_+-x_-)\lambda}\right)
            & x_-\leq \lambda \leq x_+ 
             \\            N & \lambda > x_+ \end{cases}
            \label{eq:dist_MP_cdf}
\end{equation}
Note that this distribution is more susceptible to edge effects than the semicircle distribution, especially in the $\gamma=1$ limit, and the caveat that we noted for Gaussian ensembles holds here even stronger.

\subsubsection{Jacobi ensembles}
Similar to the Laguerre ensembles, the shape of the limiting distribution for the Jacobi ensembles depends on how the parameters $a$ and $b$ scale in the $N\to\infty$ limit.

The relevant parameters are
\begin{equation}
 A = \frac2\beta \frac{a}{N}\,,\qquad B = \frac2\beta\frac{b}{N}
\end{equation}

The limiting density for the Jacobi ensembles (sometimes called the Wachter law) is
\begin{equation} \label{eq:dist_rho_j}
    \rho_\text{J}(\lambda;A,B) = \frac{A+B+2}{2 \pi} \frac{\sqrt{(x_+-\lambda)(\lambda-x_-)}}{(1-\lambda^2)}
\end{equation}
with
\begin{equation}
    x_\pm = \frac{A^2-B^2 \pm 4\sqrt{(A+1)(B+1)(A+B+1)}}{(A+B+2)^2}
\end{equation}
where the support of the distribution is in the range $-1 \leq x_- \leq \lambda \leq x_+ \leq 1$. The lower limit $x_-$ approaches $-1$ when $A\to0$, while $B\to0$ sends $x_+\to 1$.

Unfolding of the spectra of the Jacobi ensembles may be performed using the CDF obtained from integrating~\eqref{eq:dist_rho_j}. The latter can be written in a compact form as:
\begin{equation}
    I_\text{J}(\lambda;A,B) = N\times
    \begin{cases} 0 & \lambda < x_- \\
             \frac{A+B+2}{\pi} \arctan t(\lambda) - \frac{A}{\pi}\arctan (c_A t(\lambda)) - \frac{B}{\pi}\arctan(c_B t(\lambda))
            & x_-\leq \lambda \leq x_+ 
             \\            1 & \lambda > x_+ \end{cases}
            \label{eq:dist_rho_J_cdf}
\end{equation}
where
\begin{equation}
    t(\lambda) = \sqrt{\frac{\lambda-x_-}{x_+-\lambda}}\,, \qquad c_A = \sqrt{\frac{1+x_+}{1+x_-}}\,,\qquad c_B = \sqrt{\frac{1-x_+}{1-x_-}}
\end{equation}
A nice result is that, in the limit where both $A\to0$ and $B\to 0$, corresponding to $N\to\infty$ with $a$ and $b$ fixed:
\begin{equation}
    \rho_\text{J}(\lambda;A=B=0) = \frac{1}{\pi\sqrt{(1-\lambda^2)}} \,, \qquad I_\text{J}(\lambda;A=B=0) = N \left(\frac12+\frac1\pi\arcsin \lambda\right)
\end{equation}
for $\lambda\in[-1,1]$.

\subsection{Distribution of spacings}
After obtaining an unfolded spectrum of eigenvalues, which we denote $z_n$, we define the consecutive level spacings as:
\begin{equation} \label{eq:def_spacings}
    s_n = z_{n+1} - z_n
\end{equation}
The level spacings are assumed to always have unit mean:
\begin{equation}
    \langle s_n \rangle =1
\end{equation}
This will always be the case after the unfolding described above.

Then, we expect the spacings to follow the famous Wigner--Dyson surmise:
\begin{equation} p_{s}(s) = {\cal C}_{\beta} \,s^\beta \exp(-c_\beta s^2) \label{eq:pdf_s} \end{equation}
with the constants given by
\begin{equation} {\cal C}_\beta = 2\frac{[\Gamma(\frac{\beta+2}2)]^{\beta+1}}{[\Gamma(\frac{\beta+1}2)]^{\beta+2}}\,,\qquad c_\beta = \left(\frac{\Gamma(\frac{\beta+2}2)}{\Gamma(\frac{\beta+1}2)}\right)^2\end{equation}

This distribution is exact for $2\times2$ matrices from the Gaussian ensembles, but is an excellent approximation for the GBE at any $N$, and is also the expected distribution in the limit $N\to\infty$.

For the CBE, LBE and JBE, the distribution is still expected to agree, but only at large $N$. The intuition is that, in that limit, the spacing statistics are governed by the local repulsion due to the Vandermonde determinant ($\prod_{m<n}|\lambda_m-\lambda_n|^\beta$). The ``global'' details, such as the fact that CBE eigenvalues live on a circle, become negligible in determining the local statistics at large $N$.

\subsection{Distribution of spacing ratios}
Analyzing the distribution of level spacings typically requires the correct unfolding of the spectrum. For many applications, this can be a cumbersome, ambiguous, and/or imprecise process.

The spacing ratios
\begin{equation} \label{eq:def_ratios}
    r_n \equiv \frac{s_{n+1}}{s_n} = \frac{z_{n+2}-z_{n+1}}{z_{n+1}-z_n}
\end{equation}
are a more robust measure, since they depend only weakly on the unfolding transformation, and:
\begin{equation}
    r_n = \frac{I(\lambda_{n+2})-I(\lambda_{n+1})}{I(\lambda_{n+1})-I(\lambda_n)} \approx \frac{\lambda_{n+2}-\lambda_{n+1}}{\lambda_{n+1}-\lambda_n}
\end{equation}
We have only to assume that $I(\lambda)$ is a slowly varying function relative to the scale of level spacings in the original variables $\lambda_n$.

The equivalent of the Wigner--Dyson surmise for the spacing ratios is the Atas--Bogomolny--Giraud--Roux (ABGR) surmise \cite{Atas:2013dis}:
\begin{equation} p_r(r;\beta) = {\cal N}_\beta \frac{(r+r^2)^\beta}{(1+r+r^2)^{1+\frac32\beta}} \label{eq:pdf_r}\end{equation}
with the normalization constant
\begin{equation}
    {\cal N}_\beta = \frac{3^{\frac32(1+\beta)}\Gamma(1+\frac\beta2)^2}{2\pi \Gamma(1+\beta)} 
\end{equation}
Like the Wigner--Dyson surmise, the ABGR surmise can be derived exactly for the minimal case of GBE with $N = 3$, but shown to be also the limiting distribution as $N\to\infty$ for the different types of $\beta$-ensembles.

The distribution is symmetric under the transformation $r \to 1/r$. It is convenient to work with the \textit{reduced ratios} defined by
\begin{equation} \label{eq:def_reduced_ratios}
    {\rt}_n \equiv \min(r_n,\frac1r_n)
\end{equation}
for which the PDF is simply $\tilde p_\rt(\rt;\beta) = 2 p_r(\tilde r;\beta)$, confined to the range $0\leq{\tilde r}\leq 1$.

A simple diagnostic of a given spectrum is the mean spacing ratio, which has the expectation values:
\begin{equation}
    \langle r \rangle(\beta) = \int_0^{\infty} r\,p_\beta(r)dr = \int_0^1(r+\frac1r)p_\beta(r)dr\,,\qquad \meanrt(\beta) = 2\int_0^1 r\,p_\beta(r)dr
\end{equation}
This can be evaluated numerically with ease for any $\beta$.

For some computations it is convenient to change to the variable $x$ defined by:
\begin{equation}
     \tilde r + \frac12 = \frac{\sqrt{3}}{2}\tan(\frac{2x+\pi}6) \qquad \Leftrightarrow \qquad x = 3\arctan(\frac{1+2\tilde r}{\sqrt3})-\frac\pi2
\end{equation}
With this definition $x(\tilde r)$ is a monotonic function going from $x(0)=0$ to $x(1) = \frac\pi2$. In terms of $x$, the PDF takes on an especially simple form, as
\begin{equation}
    p_x(x) = \frac{2^{2+\beta}}{3^{\frac32(1+\beta)}}{\cal N}_\beta (\sin x)^\beta
\end{equation}

Using this representation, it is possible to derive a general expression for the CDF, in terms of $x$:
\begin{equation} \label{eq:cdf_r}
    P_x(x) = 1 - \frac{2^{1+\beta}[\Gamma(1+\frac\beta2)]^2}{\pi \Gamma(1+\beta)}   {}_2 F_1\left(\frac12,\frac{1-\beta}{2},\frac32,\cos(x)^2\right)\cos(x)
\end{equation}
The CDF of $\rt$ can be computed by the substitution
$P_\rt(\rt) = P_x(x(\rt))$. The CDF of $r$ can be obtained by using the identification $r = \rt$ for $r\leq 1$ and $r = 1/\rt$ else.


In the original reference \cite{Atas:2013dis}, the residual PDF
\begin{equation}
    q_N(r) = p_N(r)- p_r(r)
\end{equation}
was evaluated numerically for the classical ensembles at $\beta = 1,2,4$. Here we define $p_{N}$ as the distribution of the spacing ratios at a given finite value of $N$, while $p_\beta(r)$ is the ABGR surmise~\eqref{eq:pdf_r}. Empirically, it was found that $q_N(r)$ can be parametrized as:
\begin{equation} \label{eq:pdf_r_residual}
    q_N(r) = \frac{C_N}{(1+r)^2}\left[\frac{1}{\left(r+\frac1r\right)^{\beta}} - \frac{c_\beta}{\left(r+\frac1r\right)^{\beta+1}}\right]
\end{equation}
with $C_N$ decreasing with $N$, and expected to vanish as $N\to\infty$.

The coefficient $c_\beta$ can be obtained from the condition $\int_0^1 q_N(r)dr = 0$. By changing the integration variable to $t =\frac{1-r}{1+r}$, one can find a closed-form expression for it at any $\beta>0$:
\begin{equation}
    c_\beta = \frac{2\beta+3}{\beta+1}\,\frac{(\beta+1)c_-(\beta)+(2\beta+1)c_+(\beta)}{(\beta+2)c_-(\beta+1)+(2\beta+3)c_+(\beta+1)}
\end{equation}
in terms of the hypergeometric functions:
\begin{equation}
    c_\pm(\beta) = {}_2 F_1(\pm\frac12,\beta+1,\beta+\frac32,-1)
\end{equation}

The issue with utilizing the correction \eqref{eq:pdf_r_residual} is that it was established only numerically, and there is no known expression for $C_N$. In principle, one can introduce $C_N$ as an additional fitting parameter, but, as we will see in our numerics later, the distribution~\eqref{eq:pdf_r} can effectively model the distributions obtained at any $N$ if one allows $\beta$ to be a continuous fitting parameter. In other words, the same finite $N$ effect of~\eqref{eq:pdf_r_residual} can be captured by having an effective value $\beta_\eff$ which is typically slightly smaller than the original $\beta$.

\subsection{Non-adjacent \texorpdfstring{$k$}{k}-spacings and \texorpdfstring{$k$}{k}-spacing ratios}
To characterize spectra beyond adjacent level spacings, one can look at the non-consecutive $k$-spacings,
\begin{equation} \label{eq:def_k_spacings}
    s^{(k)}_n \equiv z_{n+k} - z_n
\end{equation}
of which adjacent spacings are the special case $k=1$.

A generalization of the spacing ratios is given by the $k$-spacing ratios defined as:\footnote{There are other similar objects that one could define here. For example $s^{(k)}_{n+1}/s_n^{(k)}$ or $s^{(1)}_{n+k}/s^{(1)}_{n}$ define two additional, nonequivalent variables. The motivation for our choice is to have $z_{n+k}$ appear in both numerator and denominator.}
\begin{equation} \label{eq:def_k_ratios}
    r^{(k)}_n \equiv \frac{s^{(k)}_{n+k}}{s^{(k)}_n} = \frac{z_{n+2k}-z_{n+k}}{z_{n+k}-z_n}
\end{equation}

There is evidence, both numerical and analytic, that in the $N\to\infty$ limit the distribution of the $k$-spacings and $k$-ratios is of the same form as the Wigner--Dyson and ABGR surmises, \eqref{eq:pdf_s} and \eqref{eq:pdf_r}, with an effective $k$-dependent value of $\beta$ \cite{AbulMagd:1999,Tekur:2018nme}:
\begin{equation} \label{eq:beta_k}
    \beta^{(k)} = \frac{k(k+1)}2\beta + (k-1)
\end{equation}
Note that in the case of the $k$-spacings one has to normalize them to unit mean by taking $\bar s_{n}^{(k)} = s_n^{(k)}/k$ before matching with the Wigner--Dyson surmise.

In this work we verify this formula for non-integer values of $\beta$, and see that, while the distributions can still be fitted with the predicted effective $\beta$ at small $k$, there are large deviations at finite $N$ from eq.~\eqref{eq:beta_k} as $k$ is increased.

\subsection{The spectral form factor} \label{sec:sff}
The spectral form factor is a function of the eigenvalues and a conjugate ``time'' variable $\tau$ which can diagnose the distribution of eigenvalues beyond the spacing statistics. In the RMT ensembles it has a characteristic structure of a decline at early times, followed by a rising ``ramp'' region that eventually reaches a ``plateau'' at late times.

We define the spectral form factor as:
\begin{equation} \label{eq:def_sff}
    \mathrm{SFF}(\tau) = \frac1N\big\langle\sum_{n=1}^N\sum_{m=1}^N e^{2\pi i(z_m-z_n)\tau} \big\rangle
\end{equation}

Note our conventions: we usually assume the eigenvalues have been unfolded, or at least normalized to have unit mean spacing. The ``time'' variable $\tau$ is defined with a factor of $2\pi$, and we normalize the SFF with an overall factor of $1/N$. With this convention the plateau is expected to start around $\tau = 1$, and the value of the SFF at the plateau is 1.

The SFF can be decomposed into its disconnected and connected pieces. The disconnected piece is given by
\begin{equation} \label{eq:def_sff_disc}
    \mathrm{SFF}_\text{disc.}(\tau) = \frac{1}{N}|R_1(\tau)|^2
\end{equation}
with
\begin{equation} \label{eq:def_sff_R1}
    R_1(\tau) = \big\langle \sum_{n=1}^Ne^{2\pi i z_n\tau}\big\rangle
\end{equation}
This in turn is essentially the Fourier transform of the eigenvalue density function:
\begin{equation}
    R_1(\tau) = N\int d z\, \rho(z) e^{2\pi i \tau z}
\end{equation}
which we can compute for the various ensembles. We write the expression for our choice of conventions below.

For the uniform density~\eqref{eq:uniform_density} it is:\footnote{In this and following expressions we omit irrelevant phases from the Fourier transform.}
\begin{equation}
    R_1^\text{(uni.)}(\tau) = \frac{\sin(N\pi\tau)}{\pi\tau}
\end{equation}
For the semicircle density~\eqref{eq:dist_semicircle}  it is given by a Bessel function:
\begin{equation}
    R_1^\text{(s.c.)}(\tau) =N \frac{J_1(4\pi \sqrt{N} \tau)}{2\pi \sqrt{N} \tau}
\end{equation}
For the Marchenko--Pastur density~\eqref{eq:dist_MP} the general expression is more cumbersome and we do not write it here, but in the limiting case of~\eqref{eq:dist_MP_1} it is again a simple expression in terms of Bessel functions:
\begin{equation}
    R_1^\text{(MP)}(\tau) = N\left(J_0(4\pi N \tau) - i J_1(4\pi N \tau)\right)
\end{equation}
The connected piece of the SFF is then given by
\begin{equation} \label{eq:def_sff_conn}
    R_2(\tau) = \mathrm{SFF(\tau)} - \frac{1}{N} |R_1(\tau)|^2
\end{equation}
When evaluating the connected SFF, the disconnected piece can be computed using either the analytical formulas just cited, or by computing $R_1$ from data using the definition~\eqref{eq:def_sff_R1}. In our implementation we select the latter option. 

For the classical ensembles we have explicit expressions for the connected SFF \cite{Liu:2018hlr}:
\begin{align} \label{eq:sff_goe} R_2^{\mathrm{GOE}}(\tau) &= 
    \begin{cases} 2\tau - \tau\log\left(1+2\tau\right) \qquad & \tau \leq 1 \\
    2 - \tau\log\left(\frac{2\tau+1}{2\tau-1}\right) & \tau > 1 \end{cases} \\ \label{eq:sff_gue}
    R_2^{\mathrm{GUE}}(\tau) &= 
    \begin{cases} \tau  \qquad & \tau \leq 1 \\
    1 & \tau > 1 \end{cases} \\
   \label{eq:sff_gse} R_2^{\mathrm{GSE}}(\tau) &= 
    \begin{cases} \frac{\tau}{2} - \frac{\tau}{4}\log(1-\tau) \qquad & \tau \leq 1 \\
    \frac{\tau}{2} - \frac{\tau}{4}\log(\tau-1) \qquad & 1 < \tau \leq 2 \\
   1 & \tau > 2 \end{cases}
\end{align}
We wrote the formulas as if for the Gaussian ensembles, but they also hold for the corresponding Circular and Laguerre ensembles, with the best agreement expected as usual in the large $N$ limit.

In \cite{Bianchi:2024fsi}, two interpolation formulas for non-integer $\beta$ were proposed, and supported by numerical evidence. The interpolation between GOE and GUE ($1\leq\beta\leq2$) is:
\begin{equation}
 R_2^{\mathrm{OU}}(\tau) = \begin{cases} \frac{2}{\beta}\tau - \left(\frac{2}{\beta}-1\right)\tau \log(1+2\tau) & \tau < 1 \\
\frac{2}{\beta} - \left(\frac{2}{\beta}-1\right)\tau \log\left(\frac{2\tau+1}{2\tau-1}\right) & \tau \geq 1 \end{cases} \label{eq:sff_beta}
\end{equation}
The interpolation from GUE to GSE ($2\leq\beta\leq4$) is given by:
\begin{equation}
     R_2^{\mathrm{US}}(\tau) = 
    \begin{cases} \frac2\beta\tau -\frac12\left(1-\frac2\beta\right) \tau\log(1-\tau) \qquad & \tau \leq 1 \\
   \frac4\beta-1+(1-\frac2\beta)\tau-\frac12\left(1-\frac2\beta\right)\tau\log(\tau-1) \qquad & 1 < \tau \leq 2 \\
   1 & \tau > 2 \end{cases} \label{eq:sff_beta2}
\end{equation}
Note that the two formulas are simple linear interpolations between the classical ensemble results:
\begin{align}
    R_2^{OU} =& \left(\frac2\beta-1\right)R_2^\text{GOE}+\left(2-\frac{2}\beta\right) R_2^\text{GUE} \\
    R_2^{US} =&\left(\frac4\beta-1\right)R_2^\text{GUE}+\left(2-\frac4\beta\right)R_2^\text{GSE}
\end{align}
These formulas work remarkably well when compared with numerical data, and in section~\ref{sec:sff_exp} we provide some more numerical evidence for this. The main discrepancy is in the unitary to symplectic interpolation. The linear interpolation inherits a log-singularity at $\tau = 1$ from the GSE, which in fact develops only gradually when increasing $\beta$ from 2 to 4. Whether these formulas can be refined further or derived, at least in part, analytically is an open question.

\subsection{Uncorrelated (Poisson) spectra} \label{sec:Poisson}
For completeness, we list here the expressions for what we call ``Poisson spectra'', which are simply spectra consisting of uncorrelated levels. These are not related to a specific random matrix ensemble, though the Poisson spectra can be thought of as a limiting case of the $\beta$-ensemble distributions at the point $\beta = 0$, where eigenvalues become independent and no longer repel.

By our definition, the ``eigenvalues'' in Poisson spectra are $N$ i.i.d. uniform variables, i.e.,
\begin{equation} \label{eq:dist_lambda_poisson}
    \lambda_n \sim U(0,N)
\end{equation}
for $n = 1,\ldots,N$. We have chosen the normalization such that the level spacings will have unit mean.

The distribution of the level spacings in this case is the exponential distribution:
\begin{equation}
    p_s^\text{(P)}(s) = e^{-s}    
\end{equation}
and the distribution for the spacing ratios becomes
\begin{equation}
    p_r^\text{(P)} = \frac{1}{(1+r)^2}
\end{equation}

Note that these are qualitatively similar but not identical to the surmises for $\beta=0$.

Likewise, for the $k$-spacings and $k$-spacing ratios we have:
\begin{equation}
    p_{s^{(k)}}^\text{(P)}(s) = \frac{s^{k-1}}{(k-1)!}e^{-s}    
\end{equation}
and the distribution for the spacing ratios becomes
\begin{equation}
    p_{r^{(k)}}^\text{(P)} = \frac{(2k-1)!}{[(k-1)!]^2}\frac{r^{k-1}}{(1+r)^{2k}}
\end{equation}

The spectral form factor is expected to be:
\begin{equation}
    \mathrm{SFF}^\text{(P)}(\tau) = 1 +(N-1)\left(\frac{\sin(N\pi\tau)}{N\pi\tau}\right)^2 
\end{equation}
which means the connected piece should go to a flat plateau, without a ramp.

\clearpage
\section{Overview of the Python implementation} \label{sec:package_overview}
This section provides an overview of the package and the main functions used for generating and analyzing random matrix spectra. It is not intended to replace the full technical documentation, but rather meant to connect the Python implementation to the mathematical objects we introduced in the previous sections, and emphasize conceptual aspects of the implementations.

In the code snippets below, we assume that the package was installed, for instance by using:
\begin{codebox}
    pip install beta-ensembles
\end{codebox}
and is loaded as:
\begin{codebox}
import beta_ensembles as be    
\end{codebox}

The package is organized in the following modules:
\begin{table}[h!] \centering
\begin{tabular}{l|l}
Module& Scope \\ \hline

circular & Circular $\beta$-ensemble generators and auxiliary functions \\

gaussian & Gaussian $\beta$-ensemble generators and auxiliary functions \\

laguerre & Laguerre $\beta$-ensemble generators and auxiliary functions \\ 

jacobi & Jacobi $\beta$-ensemble generators and auxiliary functions \\ 

classical & Auxiliary generators for classical RMT ensembles \\ 

statistics & Tools for spectral statistics: spacings, ratios, and their distributions \\

sff & Tools to compute the spectral form factor from data and compare to predictions \\ 

plot & Plotting and visualization helper functions \\
\end{tabular}
\end{table}

\subsection{Defining and sampling from ensembles}
To define an ensemble one can instantiate an \texttt{Ensemble} object by calling one of:
\begin{table}[h!] \centering
\begin{tabular}{l|l}
Function & Definition \\ \hline

\texttt{gaussian.Ensemble} & Gaussian $\beta$-ensemble \\

\texttt{laguerre.Ensemble} & Laguerre $\beta$-ensemble \\

\texttt{jacobi.Ensemble} & Jacobi $\beta$-ensemble \\

\texttt{circular.Ensemble} & Circular $\beta$-ensemble \\ \hline

\texttt{classical.Ensemble} & Any of the 12 classical ensembles \\

\end{tabular}
\end{table}

An \texttt{Ensemble} can be created by passing it the required parameters that define the ensemble, for example, one can define:
\begin{codebox}
    gbe = be.gaussian.Ensemble(beta=2.7, N=100)
\end{codebox}
to get the GBE with $N = 100$ and $\beta = 2.7$, or
\begin{codebox}
    jbe = be.jacobi.Ensemble(beta=2.7, a = 13.2, b = 203, N=100)
\end{codebox}
to define the JBE with the listed parameters.

In addition, like all random generators in this package, an \texttt{Ensemble} may receive a random number generator (RNG) as an optional argument \texttt{rng}. This is implemented for users interested in generating exactly reproducible results, by utilizing a \texttt{numpy.random.Generator} object with a fixed seed.

After defining the ensemble, one can use the same workflow. Each Ensemble realizes the following main functions:
\begin{table}[h!] \centering
\begin{tabular}{l|l}
Function & Definition \\ \hline

\texttt{spectra} & Generate random spectra from the ensemble  \\

\texttt{matrices} & Generate random matrices from the ensemble  \\

\texttt{density} & Return the limiting density of the ensemble \\

\texttt{unfold} & Unfold the given spectra using the CDF of the ensemble \\

\end{tabular}
\end{table}

The intended workflow is that one can define an Ensemble object \texttt{ens} and then proceed to generate spectra:
\begin{codebox}
    spectra = ens.spectra(n_spectra = 1000)
\end{codebox}
and analyze them using the provided functions. One may start by unfolding
\begin{codebox}
    unfolded = ens.unfold(spectra)
\end{codebox}

The classical ensembles can be defined using \texttt{classical.Ensemble}. These are standard dense matrix constructions, implemented for completeness of the package. Defining a classical ensemble is done by passing the name of the ensemble (i.e. one of GOE, GUE, GSE, LOE, LUE, LSE, JOE, JUE, JSE, COE, CUE, or CSE), and the correct parameters for each ensemble, for example:
\begin{codebox}
    loe = be.classical.Ensemble("LOE", 200, 100)
\end{codebox}
will create an instance of the LOE with parameters $M = 200$ and $N= 100$.

Each classical ensemble can be translated to a general $\beta$ ensemble with the same parameters by calling its \texttt{beta\_ensemble} function, e.g.
\begin{codebox}
    lbe = loe.beta_ensemble()
\end{codebox}
will create a new \texttt{laguerre.Ensemble} object with $\beta = 1$, $N=100$, and parameter $\alpha$ corresponding to the chosen values of the classical parameters in this instance (see eq.~\eqref{eq:LBE_alpha}).

In addition to using \texttt{Ensemble} objects, one may call directly the spectrum or matrix generating functions from the different modules. We elaborate on their usage in the following subsections. The comments below are pertinent also to the case where one uses an \texttt{Ensemble}, since the underlying implementation is the same.

\clearpage
\subsection{Spectrum generation}
The functions used to generate random spectra are:
\begin{table}[h!] \centering
\begin{tabular}{l|l|l}
Function & Definition & Equation \\ \hline

\texttt{gaussian.spectra} & Generates GBE spectra & Eigenvalues distributed as~\eqref{eq:dist_lambda_GBE} \\

\texttt{laguerre.spectra} & Generates LBE spectra  & Eigenvalues distributed as~\eqref{eq:dist_lambda_LBE} \\

\texttt{jacobi.spectra} & Generates JBE spectra  & Eigenvalues distributed as~\eqref{eq:dist_lambda_JBE} \\

\texttt{circular.spectra} & Generates CBE spectra  & Eigenvalues distributed as~\eqref{eq:dist_lambda_CBE} \\ \hline

\texttt{classical.g*e\_spectra} & Generates G*E spectra & Eq.~\eqref{eq:dist_lambda_GBE} with $\beta = 1$, 2, or 4 \\

\texttt{classical.l*e\_spectra} & Generates L*E spectra & Eq.~\eqref{eq:dist_lambda_LBE} with $\beta = 1$, 2, or 4 \\

\texttt{classical.j*e\_spectra} & Generates J*E spectra & Eq.~\eqref{eq:dist_lambda_JBE} with $\beta = 1$, 2, or 4 \\

\texttt{classical.c*e\_spectra} & Generates C*E spectra & Eq.~\eqref{eq:dist_lambda_CBE} with $\beta = 1$, 2, or 4 \\ \hline

\texttt{classical.poisson\_spectra} & Generates Poisson spectra & Eq.~\eqref{eq:dist_lambda_poisson}

\end{tabular}
\end{table}

where ``*'' in the table stands for O, U, or S.

The main functions are \texttt{gaussian.spectra}, \texttt{laguerre.spectra}, \texttt{jacobi.spectra}, and \texttt{circular.spectra} which generate spectra of the Gaussian, Laguerre, Jacobi, and Circular $\beta$-ensembles. Their arguments are the Dyson index $\texttt{beta} = \beta$,  the size of the spectrum $\texttt{N} = N$, and the parameter \texttt{n\_spectra}, which is the number of spectra to generate. The parameter $\beta$ can be any real positive number. The LBE generator has the additional parameter $\texttt{alpha} = \alpha > -1$, and the JBE one has $\texttt{a} = a > -1$ and $\texttt{b}=b>-1$.

The generating functions for spectra of the twelve classical ensembles (orthogonal, unitary, and symplectic for each of the four families) in the \texttt{classical} module are provided only for completeness. For generating spectra from these ensembles it is always more efficient to use the algorithms for generic $\beta$ and set $\beta$ to 1, 2, or 4.

Lastly, we define the function \texttt{poisson\_spectra} with the arguments \texttt{N} and \texttt{n\_spectra} which generates spectra with Poisson statistics.

All the spectrum generating functions can be supplied with a seeded RNG for exactly reproducible results.

\paragraph{GBE spectra:} As a first concrete example, we can generate GBE spectra using:
\begin{codebox}
spectra = be.gaussian.spectra(beta, N, n_spectra)
\end{codebox}
\noindent This returns a \texttt{numpy} array of shape \texttt{(n\_spectra,N)}, representing \texttt{n\_spectra} different GBE spectra. Each row \texttt{spectra[j,:]} is a spectrum of dimension $N = \texttt{N}$, containing the real eigenvalues $\{\lambda^{(j)}_1,\lambda_2^{(j)},\ldots,\lambda_{N}^{(j)}\}$ distributed according to the GBE distribution of eq.~\eqref{eq:dist_lambda_GBE} with parameter $\beta = \texttt{beta}$.

By default the eigenvalues are sorted. In our convention eigenvalues obey (at large $N$) the semicircle distribution with radius $R = 2\sqrt{N}$, independently of the value of $\beta$.

If one wants to unfold the spectra generated by \texttt{gaussian.spectra}, one can use:
\begin{codebox}
    unfolded = be.gaussian.unfold(spectra)
\end{codebox}
\noindent which applies the formula~\eqref{eq:dist_semicircle_cdf} to each element. By default, it assumes the radius is $R = 2\sqrt{N}$. The unfolded eigenvalues will be uniformly distributed in the range $[0,N]$ with unit mean spacing, such that they can be compared directly to universal RMT statistics. There is an important caveat to this procedure: it typically will introduce edge effects from eigenvalues near the endpoints of the distribution, which may or may not be negligible.

\paragraph{Circular ensemble spectra:} While the usage between the GBE and CBE cases is made to be as close as possible, there are some differences following from the different nature of the matrices in the ensembles. The CBE is an ensemble of unitary matrices and their eigenvalues are always on the unit circle $\lambda_n = e^{i\phi_n}$. It is often convenient to work directly with the eigenphases $\phi_n$.

Our function \texttt{circular.spectra} has the additional arguments $\texttt{return\_phases}$ and $\texttt{normalize}$, which are both \texttt{True} by default. The former tells the function to return the eigenphases rather than the complex eigenvalues. The eigenphases will be uniformly distributed in the range $(-\pi,\pi]$. With the option \texttt{normalize=True} they will be shifted and rescaled by the factor $N/2\pi$ such that the spectra returned will consist of real values uniformly distributed in the range $(0,N]$, requiring no further unfolding. 

It is sufficient to run:
\begin{codebox}
spectra = be.circular.spectra(beta, N, n_spectra)
\end{codebox}
\noindent to generate a set of sorted real eigenphases, normalized to have unit mean spacings, which can then be analyzed without further processing.

Users interested in analyzing the complex eigenvalues (for instance, to verify that they are indeed all on the unit circle), can specify \texttt{return\_phases=False} when calling \texttt{circular.spectra}.

\paragraph{A performance trade-off:} For many bulk spectral statistics, GBE and CBE ensembles exhibit the same large-$N$ limiting behavior. When generating spectra for numerical experimentation, one may choose to use either ensemble, but they involve different computational trade-offs.

GBE spectra can be generated faster, thanks to the tridiagonal construction. In this basis, we can construct the diagonals of the matrix in ${\cal O}(N)$ steps, and pass them to the dedicated solver \texttt{scipy.linalg.eigvalsh\_tridiagonal}, which computes the eigenvalues of symmetric tridiagonal matrices in ${\cal O}(N^2)$ time. This allows for efficient generation of spectra at large $N$. The downside is the necessity of unfolding the spectrum, which typically introduces edge effects.

By using CBE, one can avoid unfolding altogether. The downside is that the algorithm to generate CBE spectra is less efficient at large $N$. The elements of the random matrix are still generated in ${\cal O}(N)$ steps, but the resulting CMV matrix has a structured five-diagonal form. While in principle tools may be written to exploit the structure of this matrix, for this implementation we use standard dense matrix eigensolvers, which take ${\cal O}(N^3)$ time. Consequently the CBE generator becomes significantly slower than its GBE counterpart for $N \sim 1000$.

Note that the classical ensemble generators all use dense $N\times N$ matrices which take ${\cal O}(N^3)$ time and should be avoided for this reason.\footnote{In the case of the symplectic ensembles, the matrices are in fact realized using $2N \times 2N$ complex matrices, so they are noticeably slower than their orthogonal and unitary counterparts, whereas the performance for the $\beta$-ensemble algorithms is largely independent of $\beta$.}

\paragraph{Laguerre and Jacobi ensemble spectra:} The basic usage in generating LBE spectra is:
\begin{codebox}
    spectra = be.laguerre.spectra(beta, alpha, N, n_spectra)
\end{codebox}

Unfolding is implemented in the function
\begin{codebox}
    unfolded = be.laguerre.unfold(spectra, beta, alpha, N)
\end{codebox}
which uses the Marchenko--Pastur formula, with parameters calculated using the input values of $\beta$, $\alpha$, and $N$.

If one wishes to compare to the classical ensembles, one should be aware that we defined the generic LBE using different parameters from the classical case, but they can be mapped to each other easily. In the LOE, LUE, and LSE cases we have the integer parameter $M$ instead of $\alpha$. One can use the helper function \texttt{laguerre.alpha}, using eq.~\eqref{eq:LBE_alpha}. Namely, if we run
\begin{codebox}
spectra_LOE = be.classical.loe_spectra(M, N, n_spectra)

beta = 1
alpha = be.laguerre.alpha(beta, M, N)
spectra_LBE = be.laguerre.spectra(beta, alpha, N, n_spectra)
\end{codebox}
\noindent then \texttt{spectra\_LOE} and \texttt{spectra\_LBE} should both follow the same statistics.

Usage of Jacobi ensembles is nearly identical, with the exception that there are now two parameters $a$ and $b$, corresponding to the classical $M_1$ and $M_2$. The mapping is realized in the helper \texttt{jacobi.ab(beta, M1, M2, N)}.

We note again that one should avoid the classical generators if the purpose is anything other than comparison of classical to general $\beta$-ensembles. The functions \texttt{laguerre.spectra} and \texttt{jacobi.spectra}, like their GBE equivalent, utilize optimized algorithms relying on symmetric real tridiagonal matrices, achieving ${\cal O}(N^2)$ performance.

To avoid passing the same parameters manually to each function as in the examples above, one can define an \texttt{Ensemble} object, as described in the previous section, and use its functions to generate spectra, and unfold them with the ensemble's density function.

\clearpage
\subsection{Random matrix generation}
In addition to the main functions to generate spectra, we have the functions which generate the random matrices (``*'' = O, U, or S):
\begin{table}[h!] \centering
\begin{tabular}{l|l|l}
Function & Definition & Equation \\ \hline

\texttt{gaussian.matrices} & Generates matrices with GBE spectra  & Eq.~\eqref{eq:M_beta_GBE} \\

\texttt{jacobi.matrices} & Generates matrices with JBE spectra  & Eq.~\eqref{eq:M_beta_JBE} \\

\texttt{laguerre.matrices} & Generates matrices with LBE spectra  & Eq.~\eqref{eq:M_beta_LBE} \\

\texttt{circular.matrices} & Generates matrices with CBE spectra  & Eq.~\eqref{eq:M_beta_CBE} \\ \hline

\texttt{classical.g*e\_matrices} & Generates dense G*E matrices & - \\

\texttt{classical.l*e\_matrices} & Generates dense L*E (Wishart) matrices & - \\

\texttt{classical.j*e\_matrices} & Generates dense J*E (MANOVA) matrices & - \\

\texttt{classical.c*e\_matrices} & Generates dense C*E matrices & - \\

\end{tabular}
\end{table}

These functions are included to enable inspection of the random matrices generated by our algorithms. The package documentation explains their usage. Here we emphasize only that there is an important distinction between setting $\beta = 1$, 2, or 4 in the generic functions and using the dedicated classical ensemble functions. As one example, if we define
\begin{codebox}
M_1 = be.classical.gue_matrices(N, n_matrices)
M_2 = be.gaussian.matrices(2, N, n_matrices)
\end{codebox}
\noindent then \texttt{M\_1} and \texttt{M\_2} will both contain $N\times N$ matrices with eigenvalues distributed as in the GUE. However, while the matrices in \texttt{M\_1} will be generic complex-valued matrices from the GUE with all components randomly distributed, those in \texttt{M\_2} will always be given in a very specific basis of the symmetric, real, tridiagonal matrices. 

\subsection{Spectral density}
Each module implements the spectral density function for its ensemble:

\begin{table}[h!] \centering
\begin{tabular}{l|l|l}
Function & Definition & Equation \\ \hline

\texttt{gaussian.density} & Wigner semicircle distribution and CDF & Eqs.~\eqref{eq:dist_semicircle} and \eqref{eq:dist_semicircle_cdf} \\

\texttt{laguerre.density} & Marchenko-Pastur distribution and CDF & Eqs.~\eqref{eq:dist_MP} and \eqref{eq:dist_MP_cdf} \\

\texttt{jacobi.density} & Jacobi limiting density and associated CDF & Eqs.~\eqref{eq:dist_rho_j} and \eqref{eq:dist_rho_J_cdf} \\

\texttt{circular.density} & The uniform density and its CDF  & Eq.~\eqref{eq:uniform_density} \\

\end{tabular}
\end{table}

The densities of the Circular and Gaussian ensembles depend only on $N$. The densities in the Laguerre and Jacobi ensembles depend on all of their parameters, which must be provided to the \texttt{density} function to obtain correct results. Note that for the classical Laguerre/Wishart and Jacobi/MANOVA ensembles, one has to translate the parameters first.

For this reason, it is always simpler to define the \texttt{Ensemble} object, whether one works with classical or general $\beta$-ensembles, and then use its \texttt{density} function, which automatically passes the correct parameters to the appropriate module's functions.

All the \texttt{density} functions return the spectral density by default. Setting the optional argument \texttt{cumulative} to \texttt{True} will return the corresponding CDF. The CDFs in turn are used by the unfolding functions described above. Note that when unfolding we use $I(\lambda) = N \times \mathrm{CDF}(\lambda)$, as noted in section~\ref{sec:unfolding}.

\subsection{Spacing and spacing ratios}
After generating random matrix spectra -- or importing spectra from elsewhere -- one can use the following functions to compute the spacings and their ratios:
\begin{table}[h] \centering
\begin{tabular}{l|l|l}
Function & Definition & Equation \\ \hline
\texttt{spacings} & Computes the nearest-neighbor spacings & Eq.~\eqref{eq:def_spacings} \\

\texttt{ratios} & Computes spacing ratios & Eqs.~\eqref{eq:def_ratios} and~\eqref{eq:def_reduced_ratios} \\

\texttt{k\_spacings} & Computes the $k$-spacings & Eq.~\eqref{eq:def_k_spacings} \\

\texttt{k\_ratios} & Computes the $k$-spacing ratios & Eq.~\eqref{eq:def_k_ratios} \\

\end{tabular}
\end{table}

The functions \texttt{spacings} and \texttt{ratios} return the nearest-neighbor spacings and their ratios for a given spectrum or array of spectra:
\begin{codebox}
spacings = be.spacings(spectra)
ratios = be.ratios(spectra)
\end{codebox}
The functions \texttt{k\_spacings} and \texttt{k\_ratios} have the additional parameter $\texttt{k} = k$ and return the non-adjacent $k$-spacings and the $k$-spacing ratios.

The only point of usage we remark on is the following. The distribution of spacing ratios in the $\beta$-ensembles is always symmetric under the transformation $r \to \frac{1}{r}$. For this reason, we use by default the reduced spacing ratios ${\tilde r}_n$ defined in eq.~\eqref{eq:def_reduced_ratios}. All of the functions dealing with spacing ratios have an argument \texttt{reduced}, by default \texttt{True}, which can be specified to be \texttt{False} if one wants to examine the spacing ratios without the reduction to ${\tilde r}$.

\subsection{Distributions of spacings and spacing ratios}
The \texttt{statistics} module implements the probability density functions and cumulative distribution functions of spacings and spacing ratios for arbitrary $\beta$ as:
\begin{table}[h] \centering
\begin{tabular}{l|l|l}
Function & Definition & Equation \\ \hline
\texttt{statistics.pdf\_s} & The PDF of the Wigner--Dyson distribution & Eq.~\eqref{eq:pdf_s} \\

\texttt{statistics.cdf\_s} & The CDF of the Wigner--Dyson distribution & - \\

\texttt{statistics.pdf\_r} & The PDF of the ABGR distribution  & Eq. \eqref{eq:pdf_r} \\

\texttt{statistics.cdf\_r} & The CDF of the ABGR distribution & Eq.~\eqref{eq:cdf_r} \\
\end{tabular}
\end{table}

The table is self-explanatory. We also implement eq.~\eqref{eq:pdf_r_residual} as \texttt{statistics.pdf\_r\_residual}. This takes an additional parameter $\texttt{C\_N} = C_N$, which in principle can be fitted to data. At present, this function is not used elsewhere.

\subsection{Fitting to the distributions}
We implement simple fitting routines for $\beta$ in the following functions:
\begin{table}[h!] \centering
\begin{tabular}{l|l}
Function & Definition  \\ \hline

\texttt{fit\_spacings} & Fits spacings to the Wigner--Dyson distribution \\

\texttt{fit\_ratios} & Fits spacing ratios to the ABGR distribution

\end{tabular}
\end{table}

One can run them on the spacings or ratios to return the best-fit value of $\beta$:
\begin{codebox}
beta_fit_s = be.fit_spacings(spacings)
beta_fit_r = be.fit_ratios(ratios)
\end{codebox}
Note that \texttt{fit\_spacings} assumes the spacings to have unit mean before fitting, such that the only parameter is $\beta$. For the spacing ratios, the distribution is fitted always to the \textit{reduced ratios} $\rt$. Users may check separately beforehand if their data is indeed symmetric under the expected $r\to1/r$ symmetry.

\subsection{Plotting histograms and distributions}
We implement several helper functions for plotting in the \texttt{plot} module:
\begin{table}[h!] \centering
\begin{tabular}{l|l}
Function & Definition \\ \texttt{plot.hist}& Plot histogram of values \\

\texttt{plot.hist\_s} & Plot histogram of spacings \\

\texttt{plot.hist\_r} & Plot histogram of spacing ratios \\ \hline

\texttt{plot.pdf\_s} & Plot the PDF of the Wigner--Dyson distribution \\


\texttt{plot.pdf\_r} & Plot the PDF of the ABGR distribution \\


\texttt{plot.cdf\_s} & Plot the CDF of the Wigner--Dyson distribution \\


\texttt{plot.cdf\_r} & Plot the CDF of the ABGR distribution \\ \hline

\texttt{plot.density} & Plot the spectral density function of a given \texttt{Ensemble} \\

\texttt{plot.gaussian\_density} & Plot the semicircle distribution \\ 

\texttt{plot.laguerre\_density} & Plot the Marchenko--Pastur distribution \\

\texttt{plot.jacobi\_density} & Plot the Jacobi spectral density \\

\end{tabular}
\end{table}

These plot the histograms of data or the predicted distribution functions in a default style and with a default set of parameters. See documentation for usage.

\subsection{Spectral form factor}
Lastly, we have the module \texttt{sff}, which implements the following functions:
\begin{table}[h!] \centering
\begin{tabular}{l|l|l}
Function & Definition & Equation \\ \hline
\texttt{sff.compute\_sff} & Computes the SFF & Eqs.~\eqref{eq:def_sff} and~\eqref{eq:def_sff_conn} \\

\texttt{sff.predicted\_sff} & Returns the predicted SFF & Eqs.~\eqref{eq:sff_goe}--\eqref{eq:sff_beta2} \\

\end{tabular}
\end{table}

The function \texttt{compute\_sff} takes a list of spectra, and computes the average spectral form factor on that sample, on a grid of values of $\tau$. It returns both the full SFF defined in \eqref{eq:def_sff}, and the numerically evaluated connected piece. The latter is obtained by computing the disconnected piece from the sample of spectra using eqs.~\eqref{eq:def_sff_disc}-\eqref{eq:def_sff_R1} and subtracting it from the SFF.

We also provide an implementation of the interpolation of the SFF for $1\leq\beta\leq4$. To compare the empirical SFF with the predicted form returned by \texttt{predicted\_sff}, one should make sure that the time variable is normalized correctly. It is best to work always with eigenvalues normalized to have unit mean spacings.

\clearpage
\section{Demonstration: Spectral statistics for general \texorpdfstring{$\beta$}{beta}} \label{sec:demo}
In this section, we reproduce the expected spectral statistics for $\beta$ ensembles, demonstrating basic usage of our package. As a basic example, consider the following code:
\begin{codebox}
beta = 1.5
N = 100
ens = be.gaussian.Ensemble(beta, N)

n_spectra = 2000
spectra = ens.spectra(n_spectra)

unfolded = ens.unfold(spectra)
spacings = be.spacings(unfolded)
ratios = be.ratios(unfolded)
\end{codebox}


\noindent This instantiates a GBE object with a defined set of parameters $\beta = 1.5$ and $N = 100$. Then, it generates a sample of random spectra from the ensemble, with the number of spectra being 2000. It computes the unfolded spectra by using the semicircle distribution. It computes the spacings and (reduced) spacing ratios for the unfolded spectra.

Then, one can use the helper functions for plotting to verify, for example, that the spacings of the unfolded spectra match the Wigner--Dyson distribution:
\begin{codebox}
be.plot.hist_s(spacings)
be.plot.pdf_s(beta)
\end{codebox}
\noindent This plots the histogram of the spacings, and overlays it with the plot of the Wigner--Dyson surmise with the parameter $\beta$.

\begin{figure}[hp!] \centering
\includegraphics[width=0.74\textwidth]{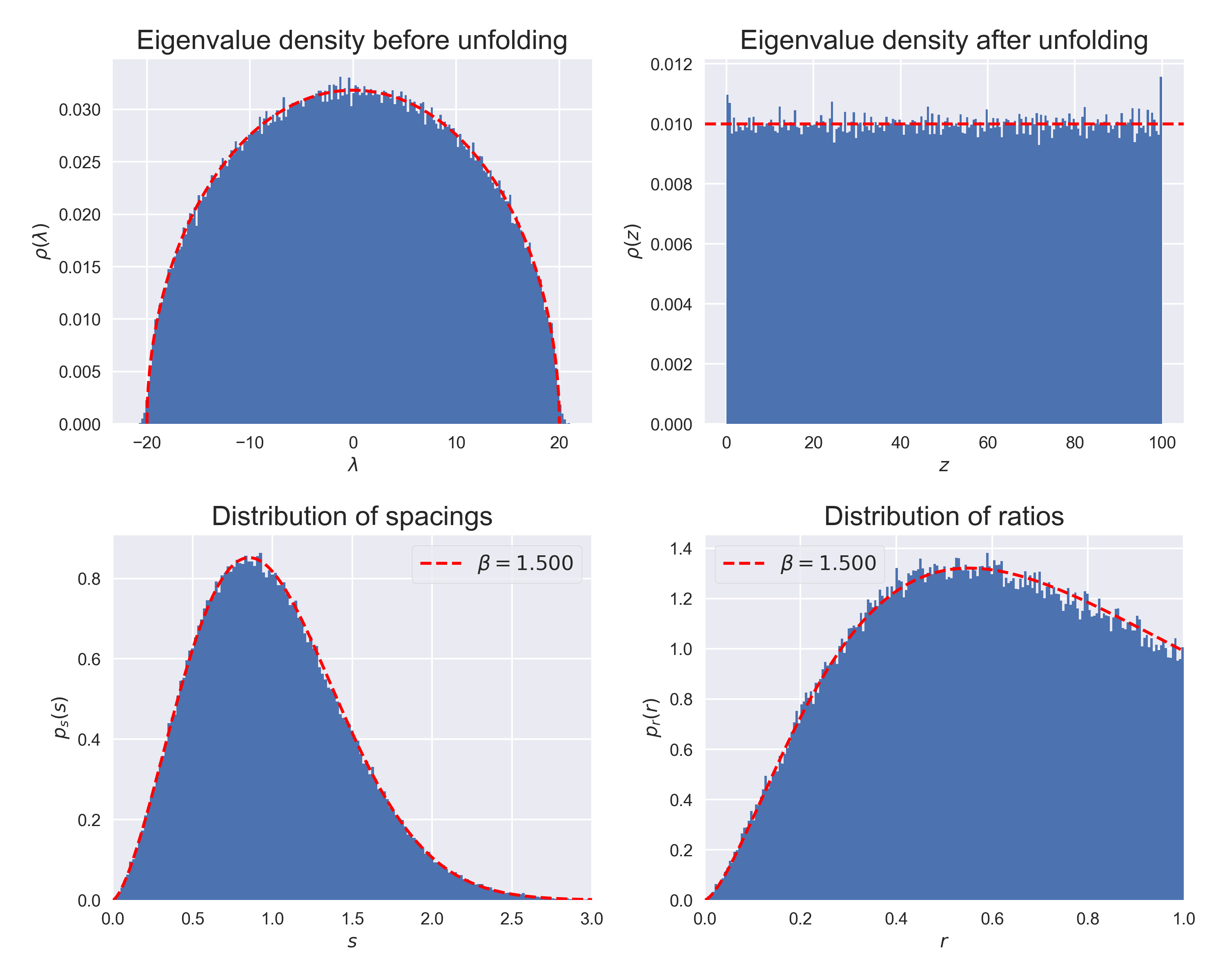}
    \caption{\label{fig:0_gbe} Results for the GBE with $\beta  = 1.5$, $N = 100$, sample of 2000 random spectra.}
\end{figure}

\begin{figure}[hp!] \centering
\includegraphics[width=0.74\textwidth]{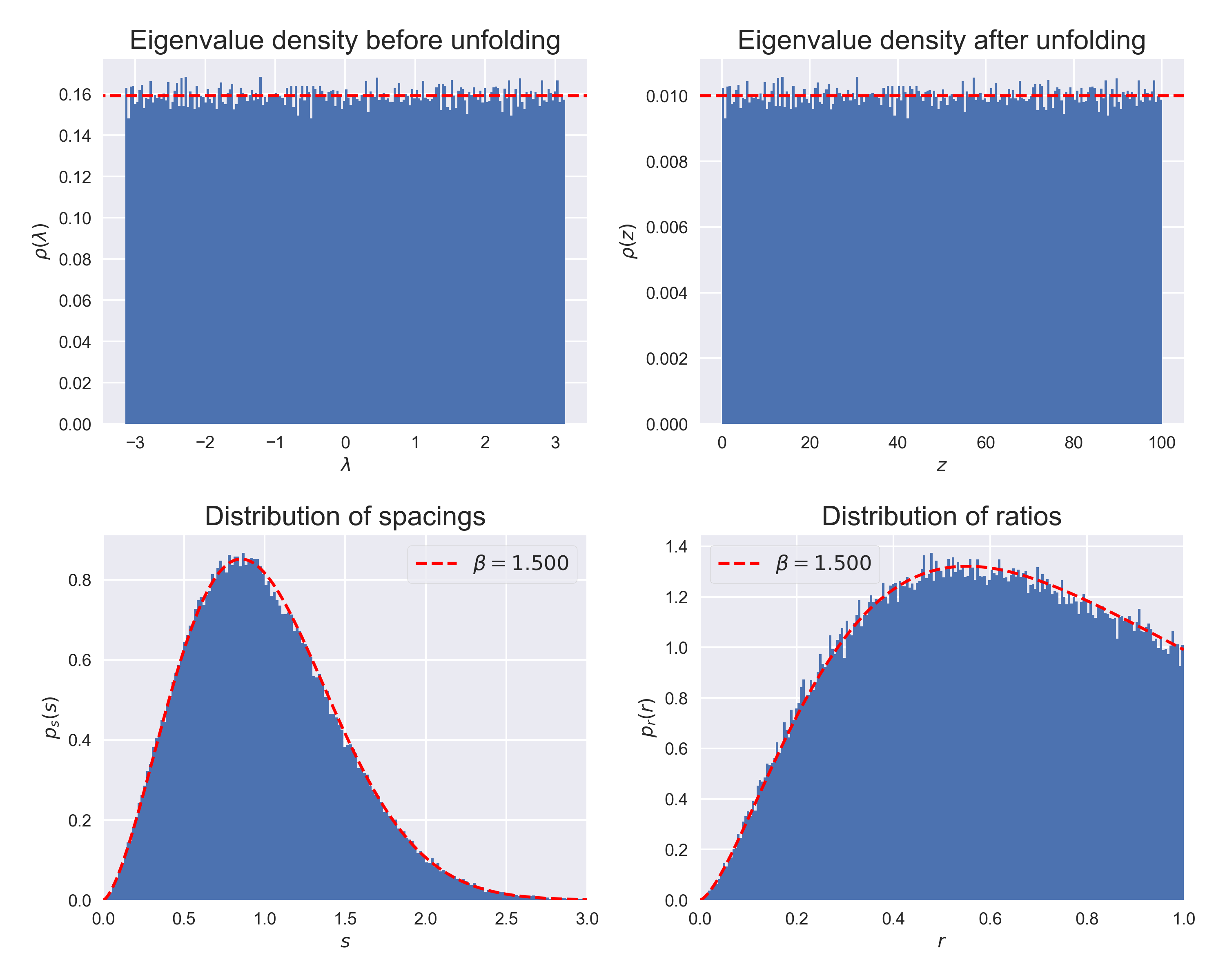}
    \caption{\label{fig:0_cbe} Results for the CBE with $\beta  = 1.5$, $N = 100$, sample of 2000 random spectra.}
\end{figure}

\begin{figure}[hp!] \centering
\includegraphics[width=0.74\textwidth]{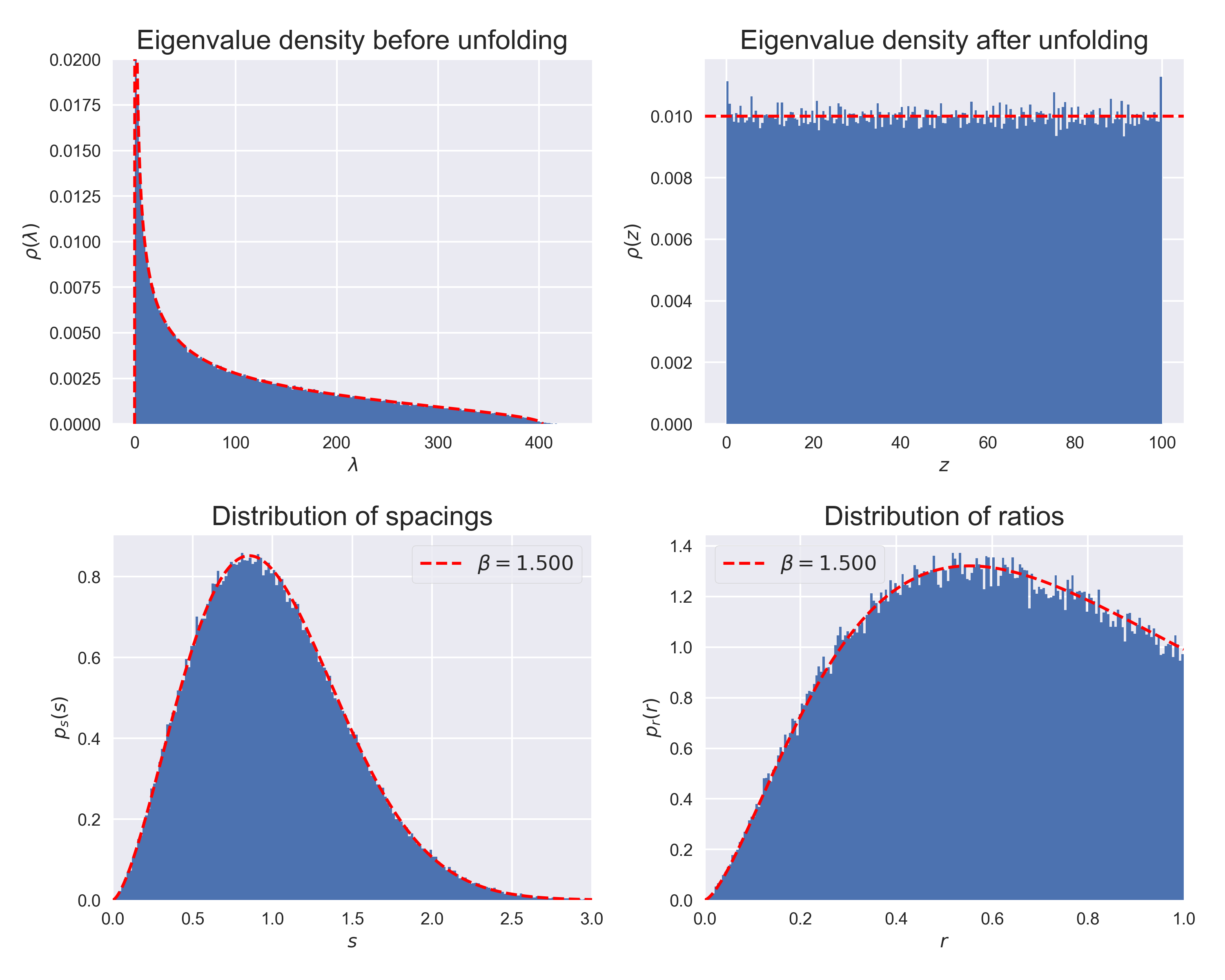}
    \caption{\label{fig:0_lbe_alpha_small} Results for the LBE with $\beta  = 1.5$, $\alpha = 1$, $N = 100$, sample of 2000 random spectra.}
\end{figure}

\begin{figure}[hp!] \centering
\includegraphics[width=0.74\textwidth]{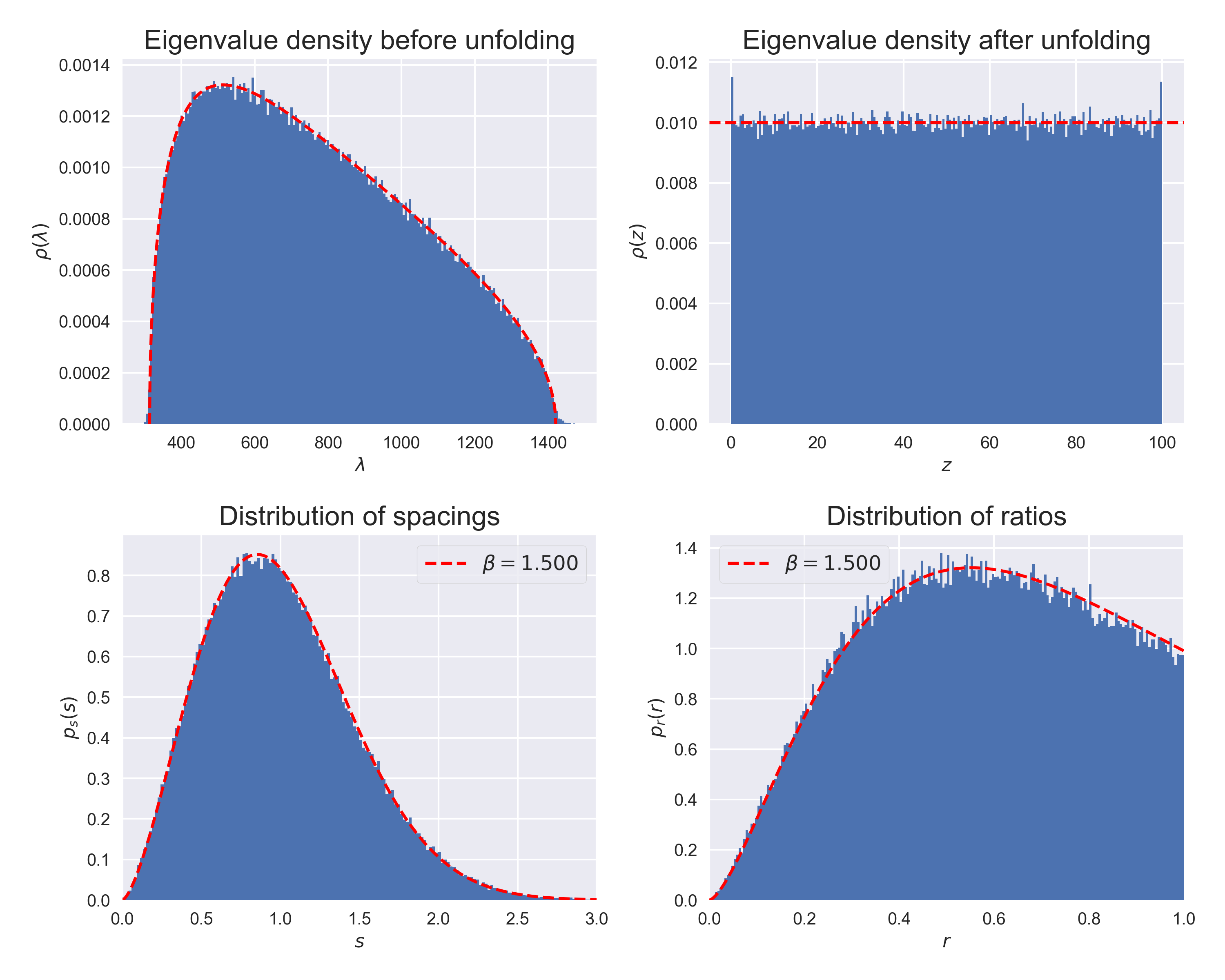}
    \caption{\label{fig:0_lbe_alpha_large} Results for the LBE with $\beta  = 1.5$, $\alpha = 500$, $N = 100$, sample of 2000 random spectra.}
\end{figure}

In figures~\ref{fig:0_gbe}--\ref{fig:0_lbe_alpha_large} below we show this basic usage example by plotting the eigenvalue densities, and distributions of spacings and spacing ratios for the three types of ensembles: Gaussian, Circular, and Laguerre.\footnote{See addendum \ref{app:jacobi} for the Jacobi ensembles.} As in the code snippet above, we set $N = 100$ and $\beta =1.5$, generating 2000 random spectra for each example, and compare them with the theoretical predictions. For the Laguerre ensembles we plot two examples, one at $\alpha = 1$ and the other at $\alpha = 500$, showing the different limiting behaviors of the Marchenko--Pastur distribution.

The data match the theoretical predictions almost perfectly. In the figures for the Gaussian and Laguerre ensembles, there are the two small spikes at $z = 0$ and $N$ of the eigenvalue density after unfolding. These are an artifact of how we handle the edge of the distribution. However, for the present figures, their effect is negligible.

These are just a few examples, sampling a small corner of the parameter space. One can also numerically explore more exotic regions instead of the familiar $1\leq\beta\leq4$. One direction is going to very large $\beta$. In this case, one can observe significant fluctuations in the eigenvalue density around the expected limiting distributions, and see that, in spite of the deviations, the unfolding still exposes the expected universal spacing statistics. Some examples are given in the repository's notebook, where the parameters $N$, $\beta$, and $\alpha$ can be freely adjusted.

\clearpage

\clearpage
\section{Applications: Exploring continuous \texorpdfstring{$\beta$}{beta}-ensembles} \label{sec:applications}
We will now utilize the package described to perform some numerical experiments, exploring various aspects of the spectral statistics of the continuous $\beta$-ensembles. Our experiments are geared toward the cases where $\beta$ is a continuous real parameter, and are largely motivated by our own work in \cite{Bianchi:2023uby,Bianchi:2024fsi}, where a non-integer $\beta$ emerged from fitting data obtained from physical scattering amplitudes.

All the results reported below can be exactly reproduced by running the example notebooks provided in the code repository. They can also be easily extended or modified to continue the studies initiated here.

\subsection{Experiment I: Continuous \texorpdfstring{$\beta$}{beta} as a fitting parameter} \label{sec:fitting_beta}
In this first experiment, we will examine $\beta$ as a continuous fitting parameter. We will draw samples of spectra from the Gaussian $\beta$-ensembles and then compare the measured distribution of spacing ratios to the ABGR surmise, the distribution~\eqref{eq:pdf_r}. We will see that the results fit well the surmise with an effective value of $\beta$ that is always slightly smaller than the input value. In the second part, we attempt to quantify the uncertainties in the fitting procedure and examine how statistical goodness of fit measures behave in this simple setup of fitting GBE spacing ratios to their expected distribution.

\subsubsection{Fitted values of \texorpdfstring{$\beta$}{beta} at fixed \texorpdfstring{$N$}{N}}
We begin with the following experiment. For each $\beta$ in $\beta_\text{in} = \{0.5,1,1.5,\ldots,6\}$, we generate two samples of $5000$ GBE spectra, one of dimensions $N = 100$ and the second with $N = 1000$. From each spectrum we keep only the 80\% bulk eigenvalues, by dismissing the $N/10$ smallest and largest values. For the remaining eigenvalues we compute the reduced spacing ratios $\tilde r_n$.
    
\begin{figure}[h!]
    \centering
    \includegraphics[width=0.48\textwidth]{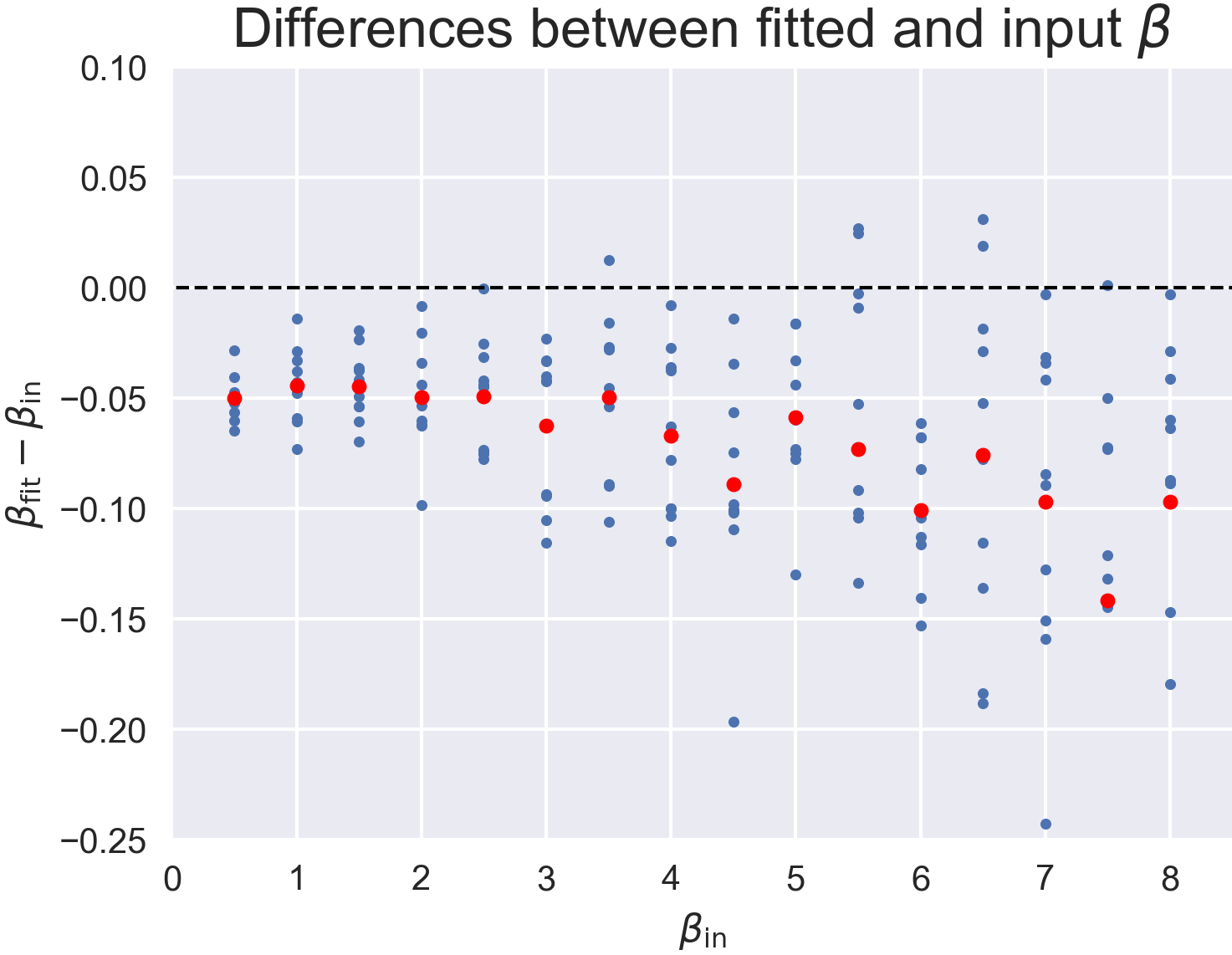}
    \includegraphics[width=0.48\textwidth]{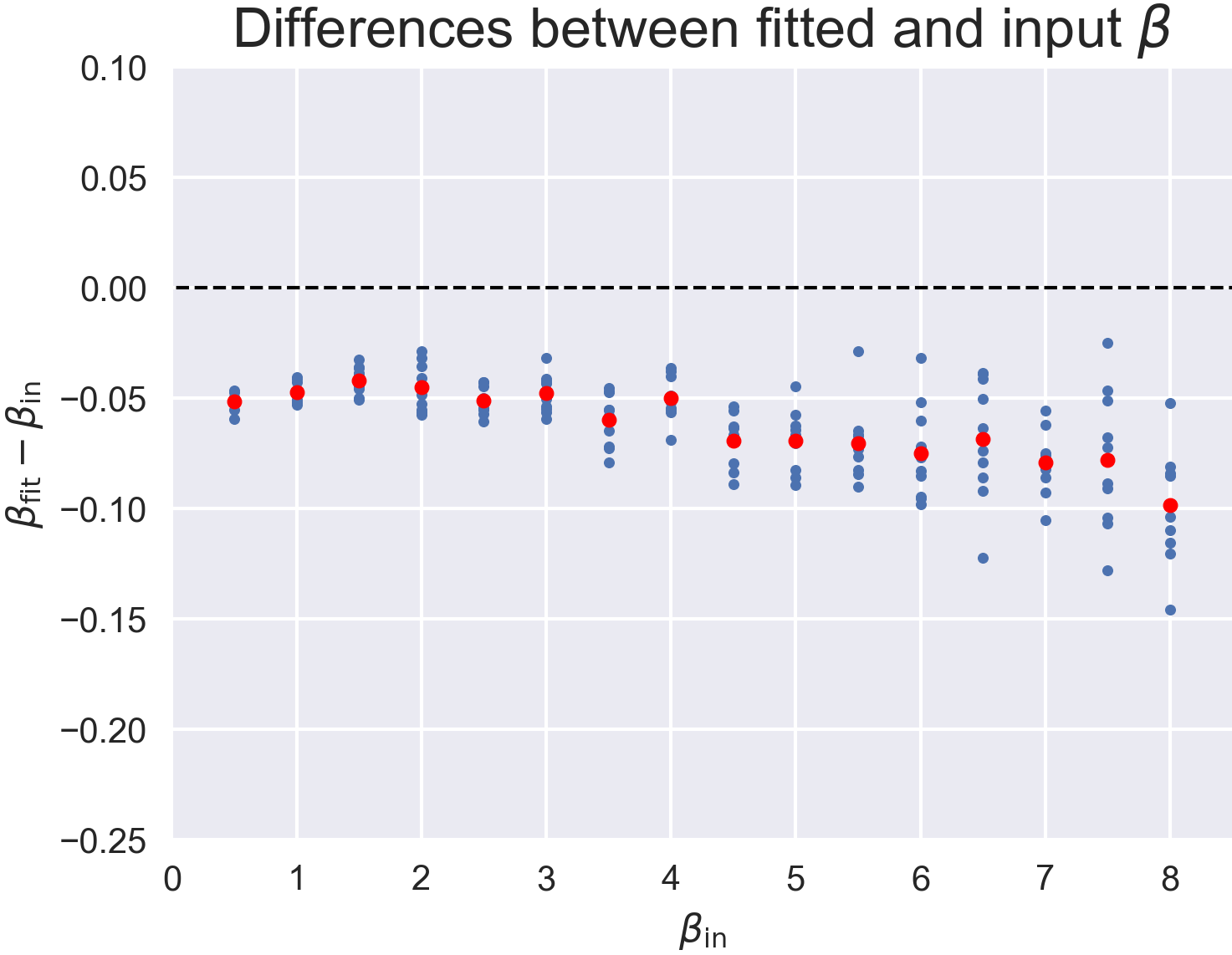} \\ 

    \includegraphics[width=0.48\textwidth]{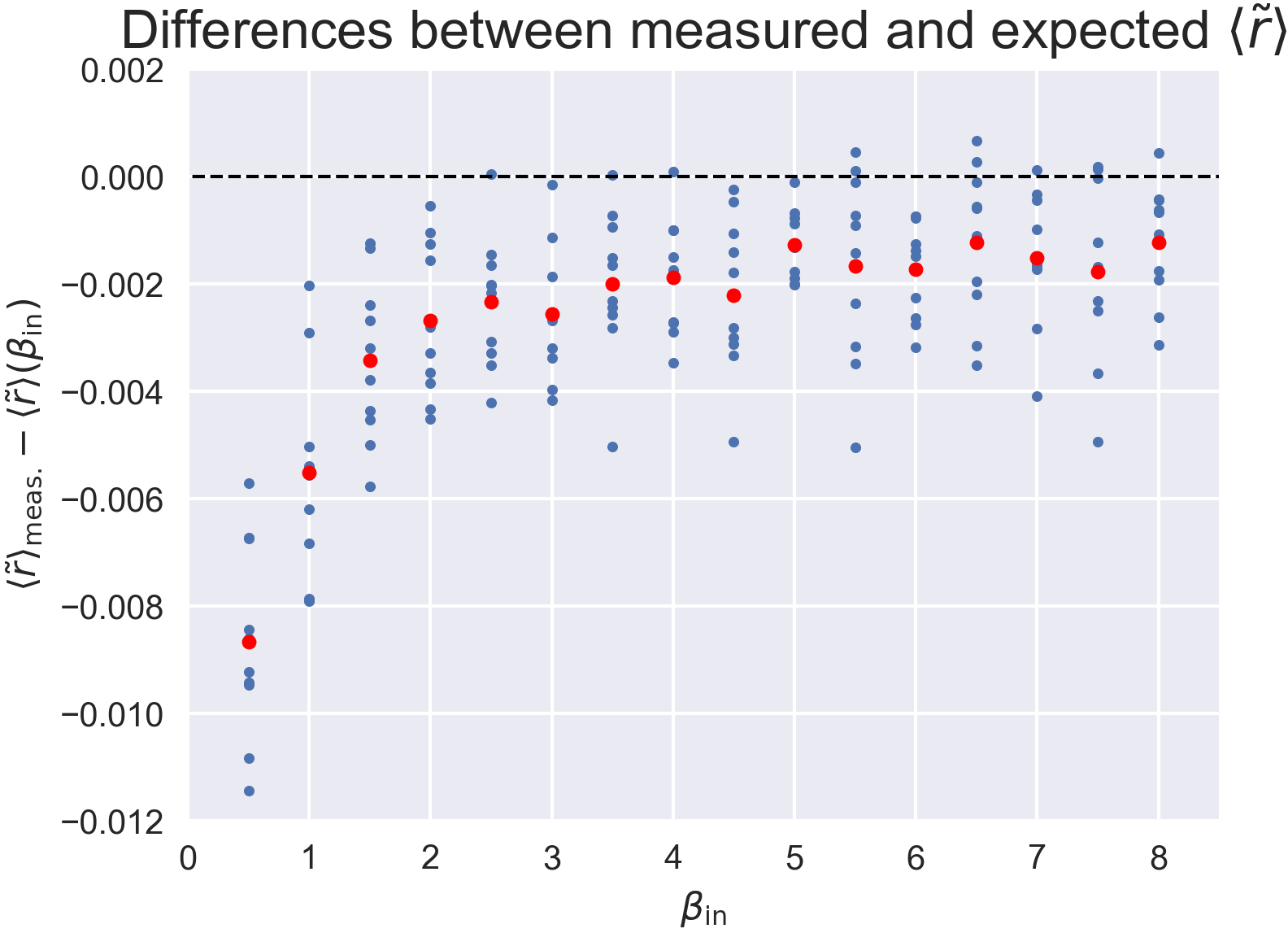}
    \includegraphics[width=0.48\textwidth]{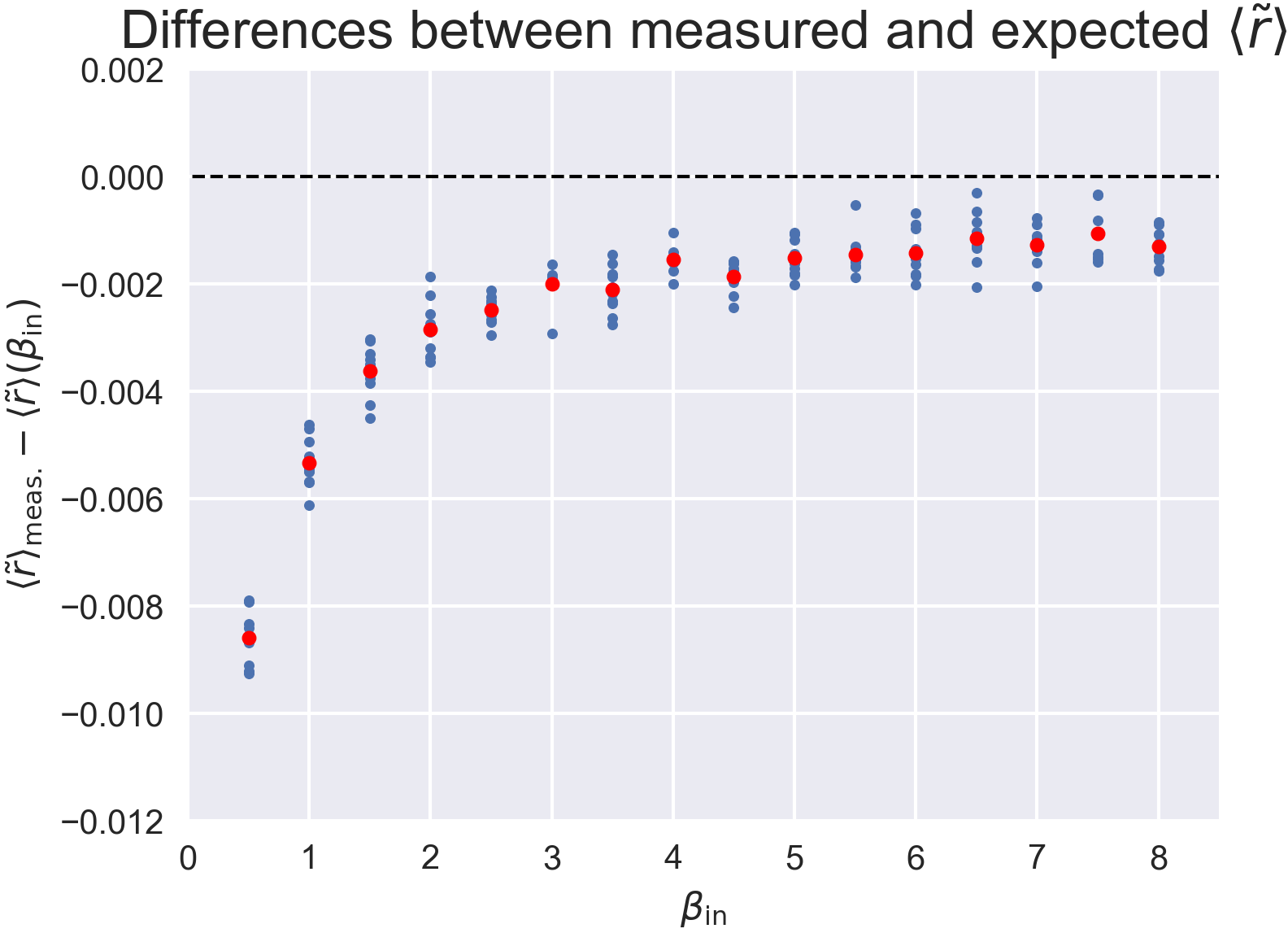}
    \caption{Top row: Deviation of best-fit $\beta$ from input value. Bottom: deviation of measured $\meanrt$ from expected value. Left column: $N = 100$, right: $N = 1000$. The blue dots are measurements on groups of 500 spectra, the red dots are the measurement on the entire sample of 5000 spectra.}
    \label{fig:beta_fit_vs_beta}
\end{figure}

Each sample is divided into 10 separate groups of 500 spectra. For each group, we fit the spacing ratios to the ABGR distribution and extract the best-fit value of $\beta$. We also extract a best-fit value for the entire sample of 5000 spectra. In addition, we compute $\meanrt$ in each group/sample to compare with the expected value. The results are plotted in figure~\ref{fig:beta_fit_vs_beta}.

There is a clear and systematic deviation: the fitted values of $\beta$ are nearly always below the input value. For the large sample size, they are always approximately 0.05 below the input value. The same phenomenon is observed in the measurements of $\meanrt$, which are completely independent of the fitting procedure.

\begin{figure}[h!]
    \centering
    \includegraphics[width=0.48\textwidth]{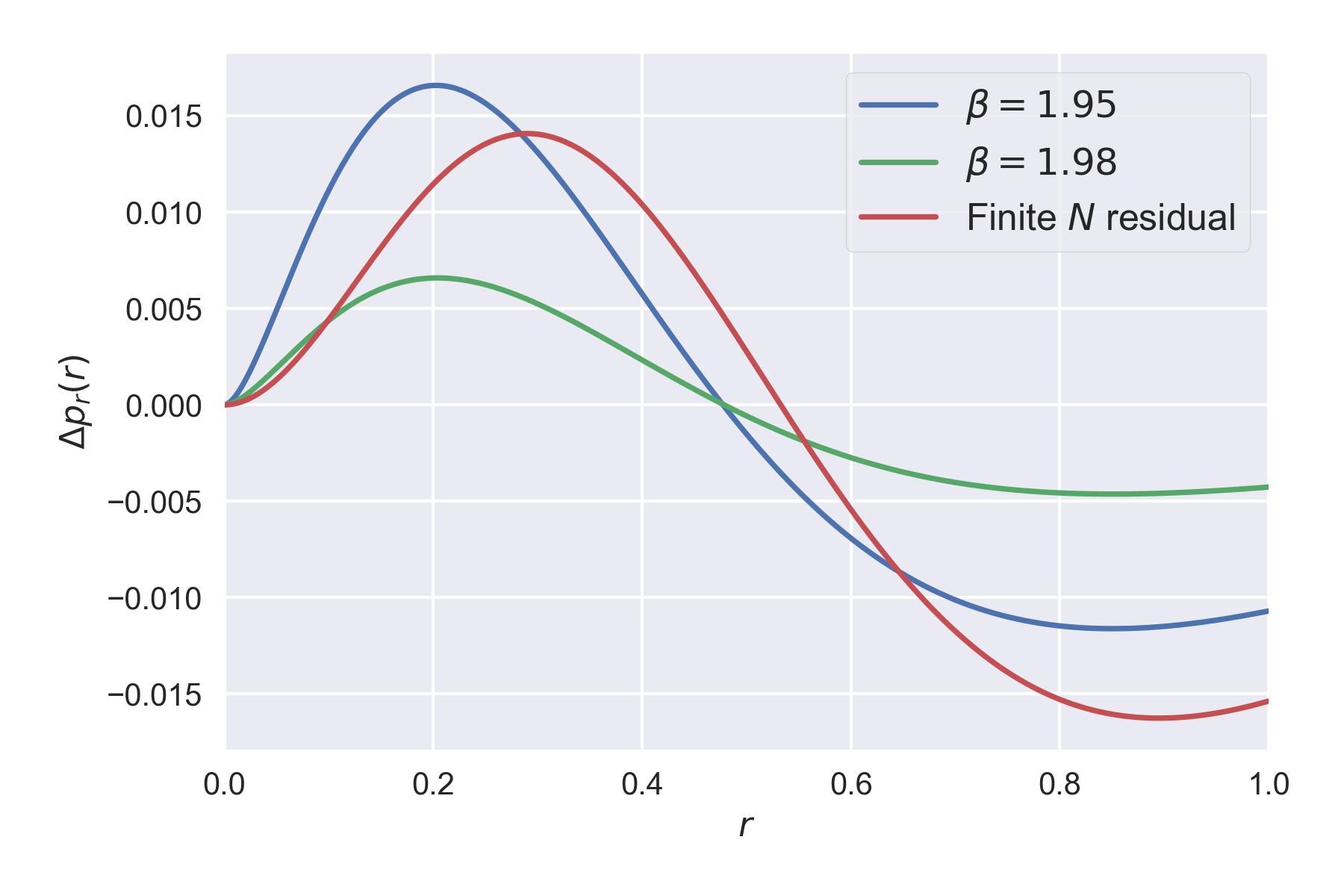}
    \caption{The residual PDF relative to the PDF of $\beta = 2$. We plot curves for two values of $\beta$ just below 2, to be compared with the ABGR residual formula~\eqref{eq:pdf_r_residual}. The behavior is qualitatively similar, though mismatched.}
    \label{fig:residual}
\end{figure}

The fact that we always get values of $\beta$ smaller than the input value is an expected effect of working at finite $N$. The ABGR surmise is exact only for $N=3$ or in the $N\to\infty$ limit, where the local spacing statistics miraculously reduce to those of the minimal $N=3$ case. We note that the finite $N$ residual PDF found empirically by ABGR is qualitatively similar to having lower $\beta$, which explains our results. For illustration, we plot in figure \ref{fig:residual} the residual PDF $q_N(r)$ defined in eq.~\eqref{eq:pdf_r_residual} for $\beta=2$, alongside the difference in the PDFs of $\beta = 1.95$ and $\beta = 2$, that is the function $p_r(r;\beta=1.95) - p_r(r;\beta=2)$. Lowering $\beta$ is qualitatively similar. The free parameter $C_N$ here was chosen as $0.6$. We have not attempted any fitting procedure in this plot. The values are simply chosen for illustration.

Note also that the plots in figure \ref{fig:beta_fit_vs_beta} are for the spacing ratios computed without unfolding. We verified that the results are essentially identical after unfolding. This can be seen in the companion notebook.

\clearpage

\clearpage

\subsubsection{Goodness of fit and uncertainties of \texorpdfstring{$\beta$}{beta}}
We saw in the previous section that the fitted value of $\beta$ is always smaller than the input value, and that there is some spread of the fitted values that becomes smaller when we take a larger sample of eigenvalues. Here we attempt a more systematic analysis to see to what extent we can trust the measurement of the parameter $\beta_\text{fit}$ and what are the uncertainties.

First, we note that the best-fit value of $\beta$ is obtained in a fitting procedure maximizing the log-likelihood function:
\begin{equation}
    L(\beta) = \sum_{i,n} \log p_\rt(\rt^{(i)}_n;\beta)\,,
\end{equation}
where $\{\tilde r^{(i)}_n\}$ is the set of all points in the fitted sample, and $p_\rt(\rt ; \beta)$ is the expected theoretical PDF at parameter $\beta$.

Near the maximum at $\beta = \beta_\text{fit}$, we can expand this function and write
\begin{equation}
    \Delta L(\beta) = L(\beta) - L(\beta_\text{fit}) \approx - \frac{(\beta-\beta_\text{fit})^2}{2\sigma^2}\,,
\end{equation}
This expansion defines $\sigma$, being the standard deviation, or error in the measurement of $\beta$.

The measure $L(\beta)$ is optimized by the fitting algorithm. We can select another, independent measure to quantify the goodness of fit (GOF). We choose the Cram\'er--von Mises (CVM) statistic.\footnote{There are alternatives. The Anderson--Darling test is a choice that gives more weight to the tails of the distribution, and is perhaps more sensitive to the precise value of $\beta$ as a result.} In the convention we use, the CVM test statistic is defined as:
\begin{equation}
    T(\beta) = \frac{1}{12 n_P} + \sum_{i=1}^{n_P} \left(\frac{2i-1}{2n_P} - P_\rt(\rt^{(i)};\beta)\right)^2
\end{equation}
The statistic essentially measures the distance between the predicted and the measured CDFs. To define it we introduced the notation $\{\rt^{(i)}\}$ for the set of all spacing ratios arranged in increasing order, and $n_P$ for the total number of data points in the set. The function $P_\rt(\rt;\beta)$ is the predicted CDF given by~\eqref{eq:cdf_r}.

Since the CVM statistic is not optimized by the fitting algorithm, we can check separately where the minimum of this function is, and how different it is from $\beta_\text{fit}$. This gives another estimate of the error in measuring $\beta$.

Our experiment now is as follows. First, we generate a random sample of 5000 spectra of dimension $N$ from the GBE with $\beta=2$. As before, we take only 80\% of the eigenvalues from each spectrum, dismissing the smallest and largest 10\%, and compute the reduced spacing ratios for them.  In addition, to see how the results depend on the size of the sample, we perform our analysis on three subgroups: the first 1000 spectra, the first 2000 spectra, and finally all 5000 spectra in the sample.

\begin{figure}[h!]
    \centering
    \includegraphics[width=0.90\textwidth]{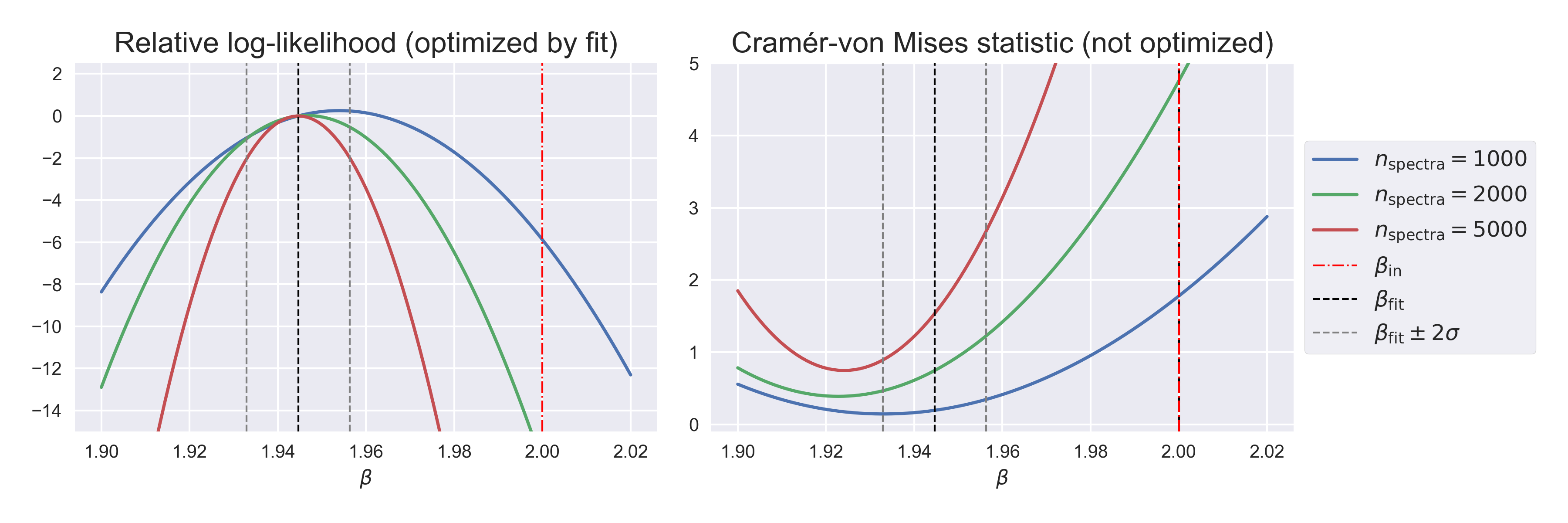} \\
    \includegraphics[width=0.90\textwidth]{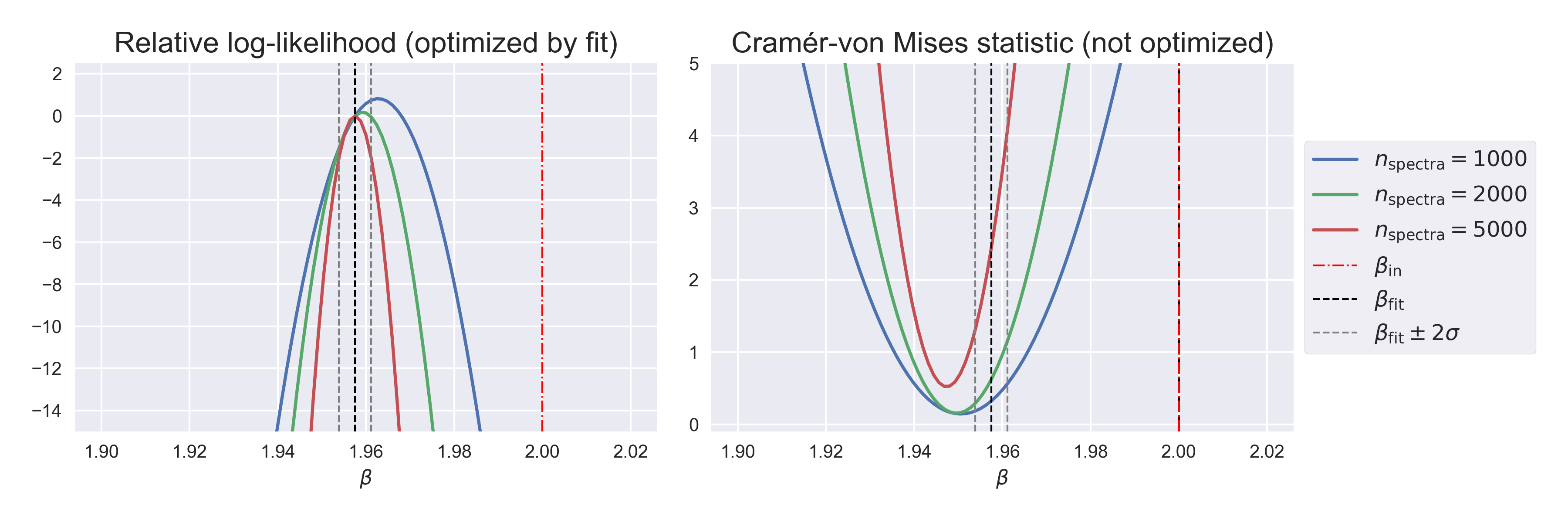} \\
    
    \caption{Log-likelihood (left) and CVM statistic (right) as functions of fitting parameter $\beta$. The log-likelihood is plotted after subtracting its value on the entire sample at $\beta=\beta_\text{fit}$. Top row: results for $N=100$. Bottom row: $N=1000$. See also results in table~\ref{tab:gof_stats}.}
    \label{fig:gof}
\end{figure}

\begin{figure}[h!]
    \centering
    \includegraphics[width=0.48\textwidth]{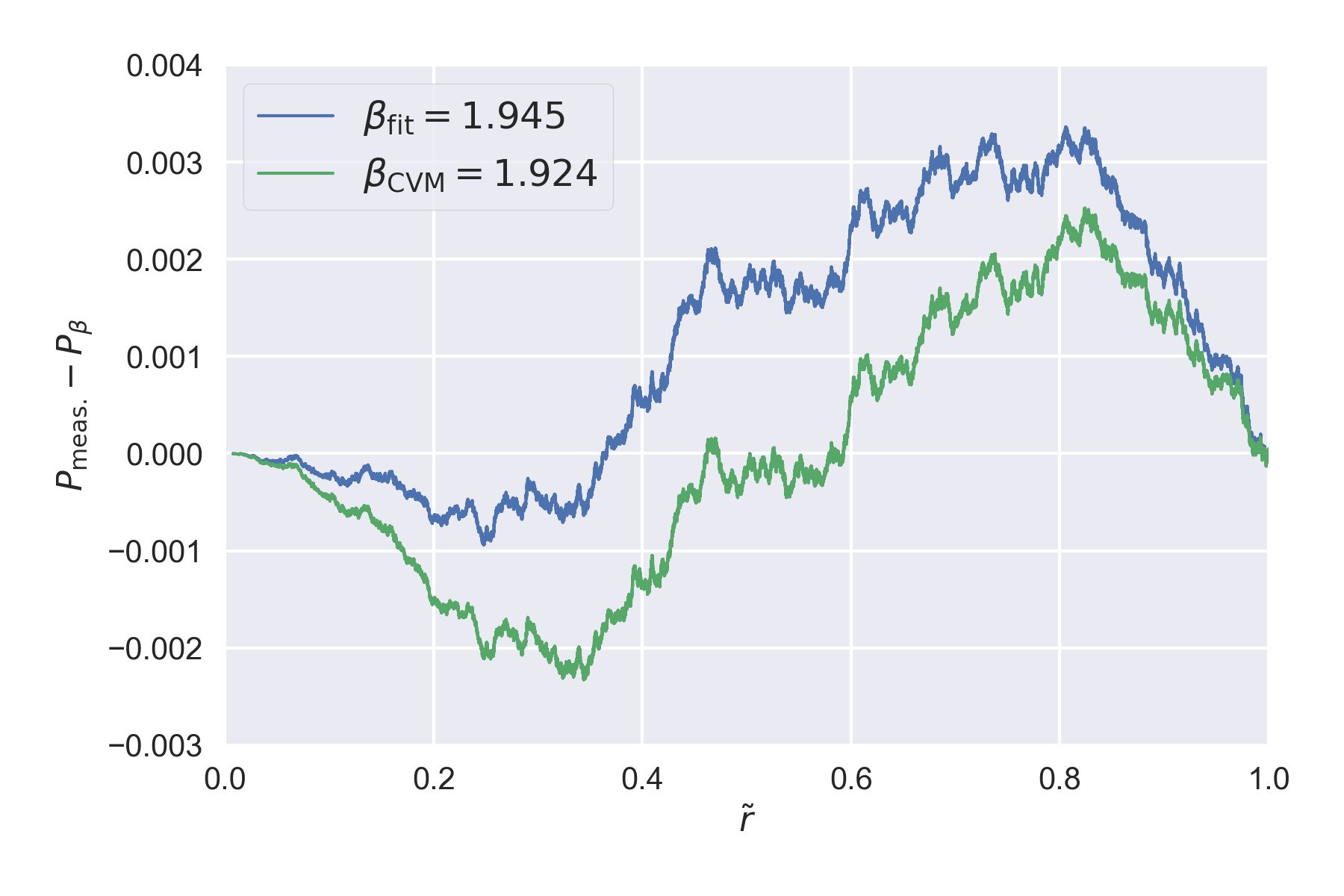}
    \includegraphics[width=0.48\textwidth]{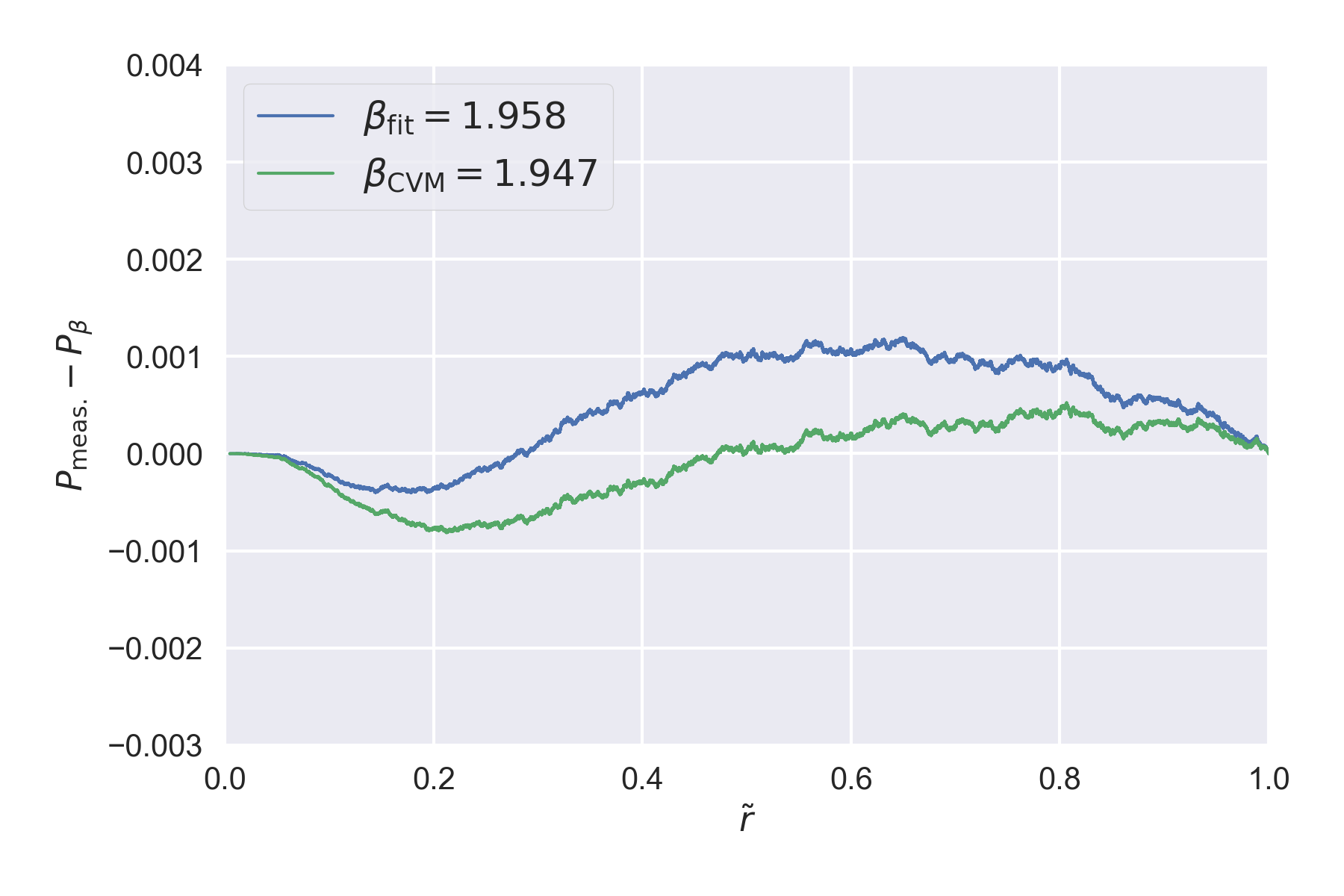} \\

    \caption{Deviation of measured from theoretical CDF. Left: result for $N=100$. Right: $N=1000$. The blue curve is for the maximum likelihood value of $\beta$, while the green curve is that which minimizes the CVM statistic.}
    \label{fig:delta_cdf}
\end{figure}

By running this experiment a few times, we observe that, not surprisingly, the two statistics typically disagree on the best value of $\beta$, especially when the sample size is large. We present here results from two typical runs, one at $N = 100$ and the other at $N = 1000$. The results are in figures \ref{fig:gof} and \ref{fig:delta_cdf} and table \ref{tab:gof_stats}. Let's note some lessons that we can learn from them.

The CVM statistic has a minimum at $\beta = \beta_\text{CVM}$ that is lower than $\beta_\text{fit} -2\sigma$, suggesting that our estimate of the error from the log-likelihood function alone was too optimistic, and depended too strongly on the fitting procedure. On the other hand, $\beta_\text{fit} = \beta_\text{in} = 2$ is strongly ruled out by both statistics. In all our runs, we found $\beta_\text{CVM} < \beta_\text{fit}$.

\clearpage

\begin{table}[th!] \centering
    \begin{tabular}{|c|c||c|c|c||c|c|c|} \hline
 $N$ & Spectra & $\beta_\text{fit}$& $T(\beta_\text{fit})$& $p$-value & $\beta_\text{CVM}$& $\min_\beta T$ & {$\max_\beta p$}\\ \hline\hline
           & 1000 & 1.954 $\pm$ 0.013&  0.305& 0.131&  1.933&   0.144& 0.410\\
 100 & 2000 & 1.947 $\pm$ 0.009& 0.811&  0.007 & 1.923&   0.388& 0.078\\
      & 5000 & 1.945 $\pm$ 0.006& 1.533& 0.000 &  1.924& 0.746& 0.010 \\ \hline\hline

           & 1000 &  1.963 $\pm$ 0.004&  0.703 &  0.013&  1.951& 0.143& 0.411\\
 1000 & 2000 &  1.959 $\pm$ 0.003 &   0.850&  0.006& 1.950&  0.155& 0.376 \\
      & 5000 &   1.958 $\pm$ 0.002& 2.455& 0.000&  1.947&   0.521& 0.035 \\ \hline
    \end{tabular}
    \caption{\label{tab:gof_stats}
    Results of our fitting experiment for $N = 100$ and $N = 1000$. The value $\beta_\text{fit}$ is that chosen to maximize the log-likelihood of the sample, with errors corresponding to 1$\sigma$. We list the values of the CVM statistic and associated $p$-value, first computed at $\beta_\text{fit}$, and second at $\beta_\text{CVM}$, the value that minimizes $T(\beta)$ on each sample.}
\end{table}

\clearpage

It is clear that with more data, the preferred range of $\beta$ (for either statistic) becomes smaller. The two statistics can agree when the total number of data points is not too large, but once more data is added, either by increasing $N$ or the number of spectra, the discrepancy grows.

Moreover, the value of the CVM statistic at $\beta = \beta_\text{fit}$ can be quite bad. In fact, for our full sample at $N=1000$, it appears that our fit is completely ruled out by the CVM test with $T \approx 2.5$. On the other hand, the minimum of $T(\beta)$ on the same sample suggests better agreement with $T \approx 0.5$ though this is still a value that suggests that the distribution is rejected. But this should not concern us too much: when we consider hundreds of thousands of data points, any systematic deviation, even very small, implies that it is statistically impossible that our sample corresponds to values drawn from the fitted distribution, even when taking into account errors in the measurement of the parameter.

We know that the systematic deviation at finite $N$ exists, as we saw in the previous section. In figure \ref{fig:delta_cdf} we now plot the deviation in terms of the CDF for our two samples, once for the value $\beta_{\text{fit}}$ and once for $\beta_\text{CVM}$. Comparing the two plots, the agreement is considerably better at $N=1000$ than at $N=100$. This is not reflected in the GOF statistics because the total number of data points is ten times larger, and more data points mean more points deviating from the theoretical prediction.

The CVM statistic attempts to minimize (the square of) the deviation plotted in the figure. It is apparent that $\beta_\text{fit}$ does a better job fitting the distribution at small values of $\tilde r$, while minimizing the CVM statistic adds deviations at small $\tilde r$, by choosing smaller $\beta$, in order to better fit at larger values. Neither option is obviously preferable.

It is important to be aware of the limitations discussed here when interpreting the fitted effective values of $\beta$ and the statistical measures of the GOF. This would be especially true when fitting physical data, which are always much messier than the idealized pure RMT spectra we considered here.

\clearpage

\subsection{Experiment II: Distributions of \texorpdfstring{$k$}{k}-spacing ratios for continuous \texorpdfstring{$\beta$}{beta}-ensembles} \label{sec:k_spacings}
In this section, we want to see whether the $k$-spacing ratios of the Gaussian $\beta$-ensembles can be fitted by the ABGR surmise, with the prediction $\beta^{(k)} = \frac{k(k+1)}{2}\beta+(k-1)$ introduced in eq.~\eqref{eq:beta_k}.

To this end, we generate GBE spectra for $N = 2^{7},2^{8},\ldots,2^{13}$, keeping the number of eigenvalues fixed at $2^{20} \sim 10^{6}$. Thus, the number of spectra decreases with increasing $N$. We choose $\beta = \pi$, for no special reason except that it is a number between $1$ and $4$ and we shouldn't always use only rational numbers.

For each $k=1,2,\ldots,13$, we compute the $k$-spacing ratios and fit their distribution to the ABGR surmise~\eqref{eq:pdf_r}, obtaining the best-fit value of $\beta$. In light of the results of the previous sections, we first fit $\beta^{(1)}$ to the full sample, averaging over all values of $N$. This gives $\beta^{(1)}\approx3.08$, slightly below the input value $\beta=\pi$, as expected. We use this fitted value in place of $\beta$ when evaluating the predicted $\beta^{(k)}$.

\begin{figure}[h!]
    \centering
    \includegraphics[width=0.96\textwidth]{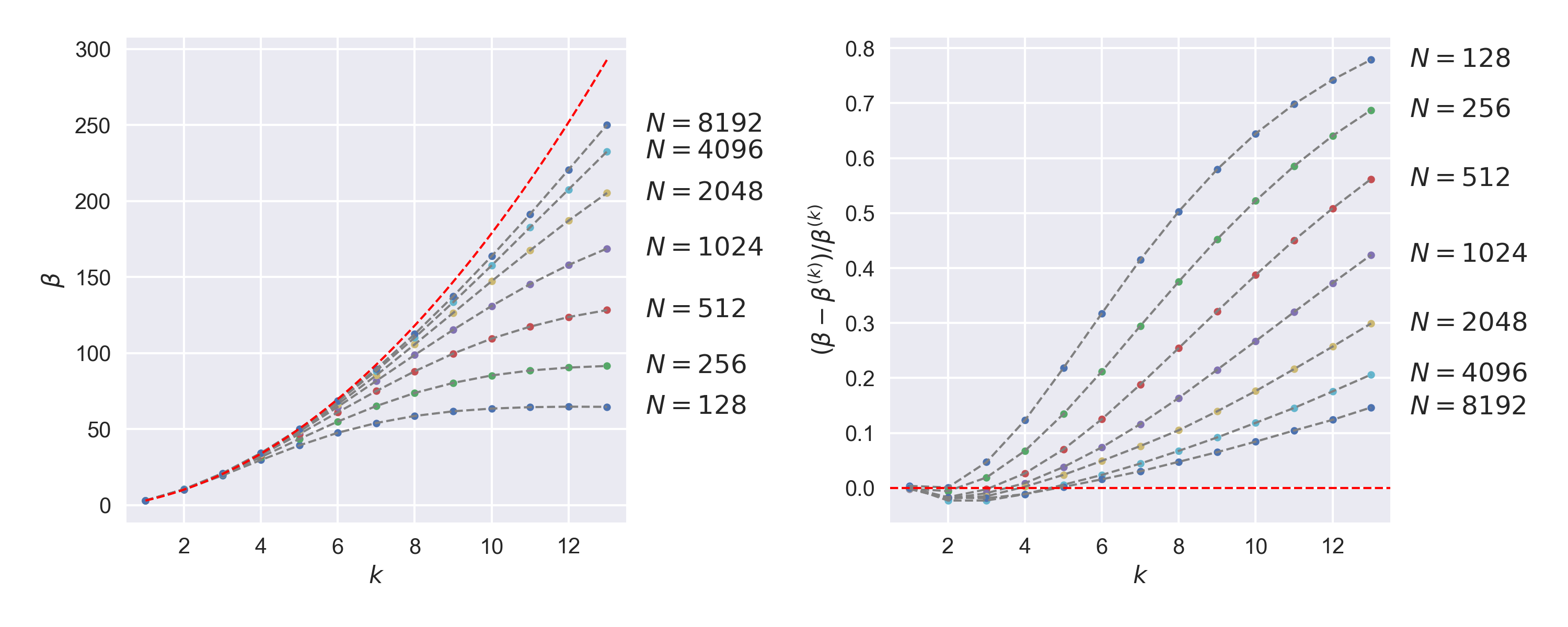} 
    \caption{The best-fit values of $\beta$ as a function of $k$ (left), with the relative deviation from predicted $\beta^{(k)}$ (right). }
    \label{fig:k_spacings_beta}
\end{figure}

The results, shown in figure~\ref{fig:k_spacings_beta}, exhibit significant finite $N$ discrepancies. The upper panels show the deviation of the fitted values of $\beta$ from the predicted $\beta^{(k)}$ as a function of $k$. Taking the $N=1024$ curve as a reference, we find a $10\%$ discrepancy at $k=7$, increasing to approximately $40\%$ at $k=13$. Increasing $N$  systematically brings the fitted values closer to the predicted curve, but the approach is very slow, perhaps logarithmic in $N$.

\begin{figure}[ht!]
    \centering
    \includegraphics[width=0.86\textwidth]{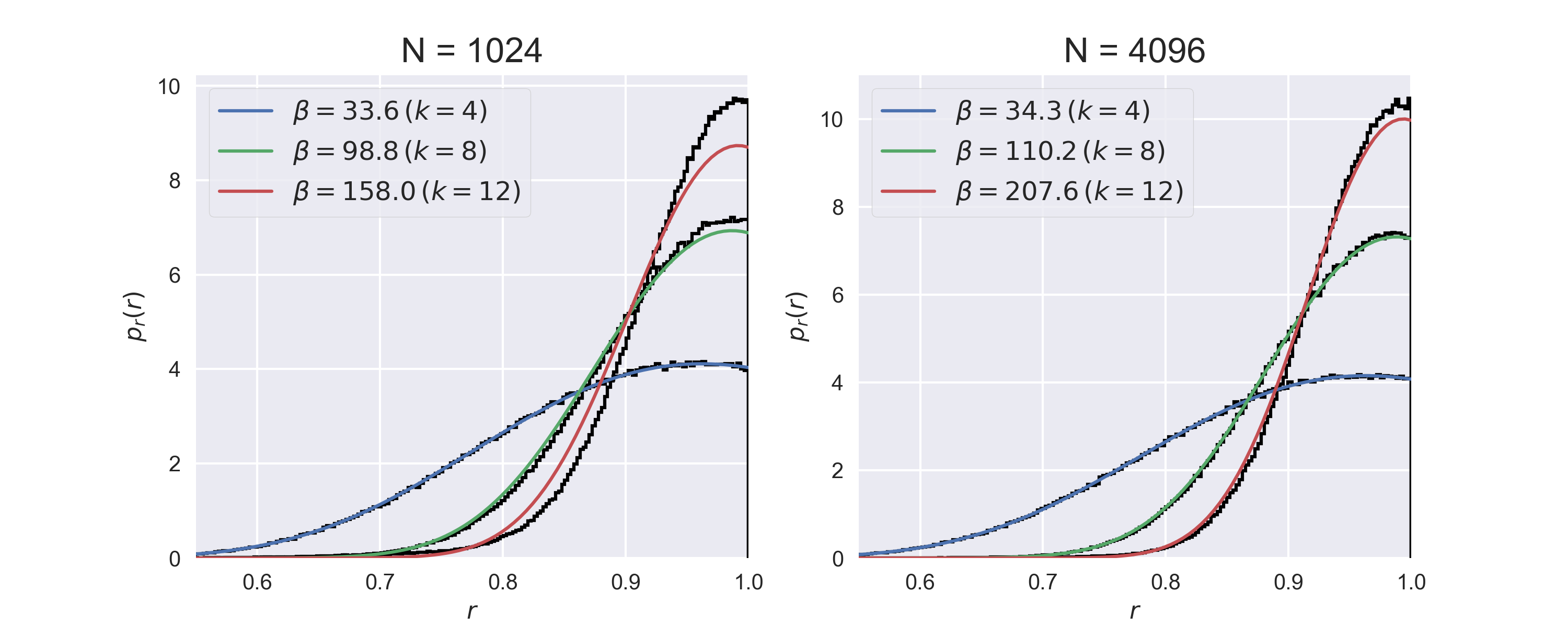}
    \caption{Histograms of $k$-spacing ratios with the best-fit $\beta$ for selected samples. The black lines represent the data, while the colored lines are the best-fitting curves.}
    \label{fig:k_spacings_hist}
\end{figure}

The histograms we plot in figure~\ref{fig:k_spacings_hist} show that, for the values of $N$ and $k$ where the deviation is large, the ABGR surmise itself provides a poor fit to the resulting distribution. Tentatively, these results suggest that the $k$-spacing ratios cannot in general be adequately described by an effective $\beta$ parameter once $k$ becomes larger than $\sim\log N$, though the scaling cannot be verified from what we do here. A more systematic investigation of this behavior can be deferred to future work. The circular ensembles would also be interesting to study in this context, given the periodic nature of their spectra.

\clearpage

\subsection{Experiment III: The spectral form factor} \label{sec:sff_exp}
The spectral form factor (SFF) allows us to probe the spectral statistics by packing the information on the eigenvalue distribution into a function of the conjugate time variable $\tau$. The SFF for generic $\beta$-ensembles has not been studied much in the literature. We mainly rely on our own work in \cite{Bianchi:2024fsi}.

The purpose here is to get some notion of how the SFF behaves at values of $\beta$ other than the usual 1, 2 and 4. An understanding of the precise results for the SFF at general $\beta$ can provide a highly non-trivial point of comparison to test data against RMT predictions, beyond the spacing distributions.

In this experiment we will use only Circular ensembles, so we won't have to worry about unfolding the spectra. The spectra of eigenphases will always have a uniform density, and after a simple rescaling all eigenvalues are in the range $[0,N]$.

We take samples of 5000 random matrices of dimension $N = 200$ at the following values of $\beta:$ 1, 1.25, 1.5, 1.75, 2, 2.5, 3.5, 4, 4.5, 6, 12, 24. For each sample we compute the SFF, as defined in section~\ref{sec:sff}. 

\begin{figure}[h!]
    \centering
    \includegraphics[width=0.96\textwidth]{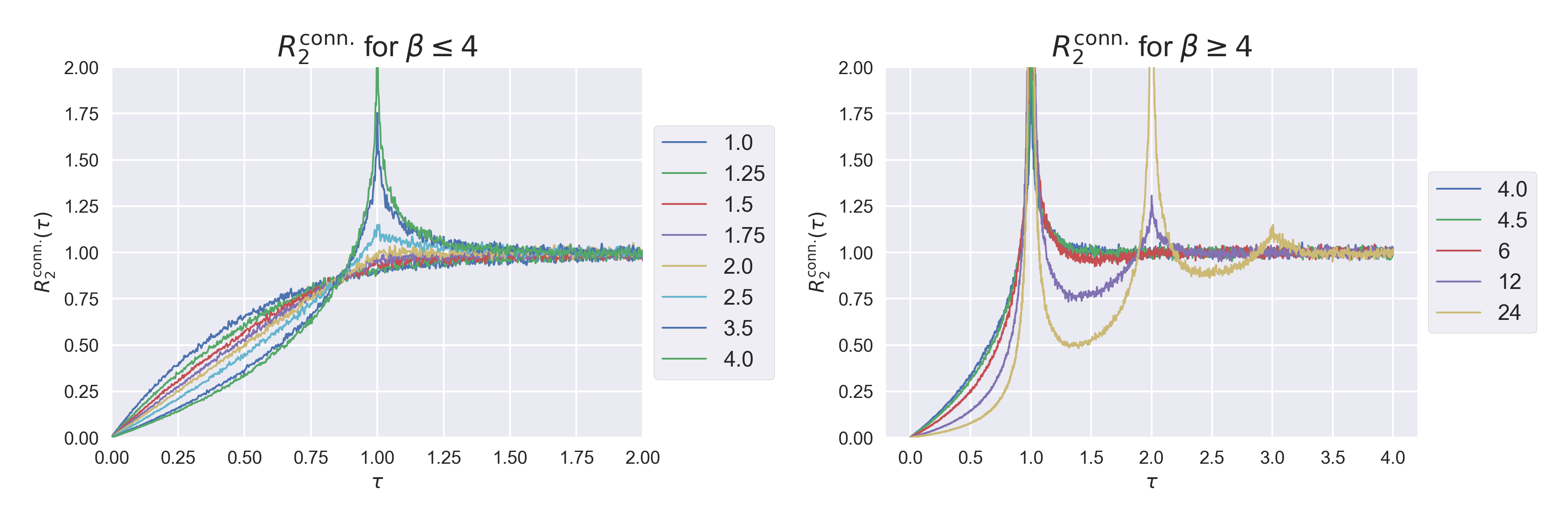} \\
    \includegraphics[width=0.96\textwidth]{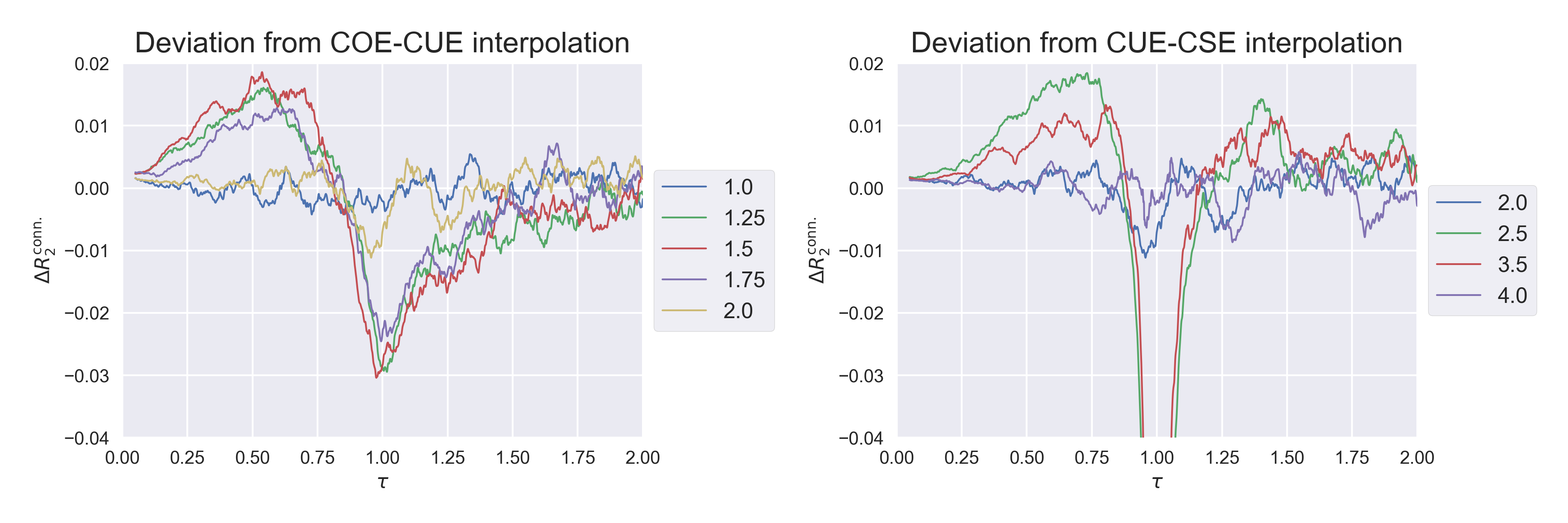}
    \caption{Top: The connected spectral form factor for the CBE at various values of $\beta$. Bottom: Deviation from the interpolation formulas.}
    \label{fig:sff_beta}
\end{figure}

We plot the results for the connected part of the SFF in figure~\ref{fig:sff_beta}. The plots show the rich structure of the SFF. There are three distinct regimes, $1\leq\beta \leq 2$, $2\leq\beta<4$ and $\beta \geq 4$, which have some qualitative differences. The classical ensembles are of course the points where the qualitative behavior changes. For $\beta$ values below 2, including the COE at $1$, the ramp of the SFF is a concave function. For $\beta = 2$ the ramp is exactly linear, with unit slope in our convention, while for $\beta > 2$ the function is convex. The main new feature of the SFF at $\beta=4$ is a logarithmic singularity at $\tau = 1$ (recall its form in eq.~\eqref{eq:sff_gse}). This develops continuously as $\beta$ is increased from 2 to 4.

In the bottom panels of the figure, we compare our results with the interpolation formulas \eqref{eq:sff_beta} and \eqref{eq:sff_beta2} by plotting the difference of the measured and ``predicted'' SFF. For the COE to CUE interpolation, the linear interpolation is an excellent approximation, with deviations of order 0.03 at most. For the CUE to CSE interpolation, there is a large deviation at $\tau=1$. The logarithmic singularity of $\beta=4$ is not compatible with the finite values of the SFF at $\tau=1$ at $\beta < 4$. Even so, elsewhere the approximation matches well, and the deviation in the neighborhood of 1 is still under 10\% for the most part.

We have attempted to refine the interpolation formulas in several ways, but have not been successful. The simple linear interpolation works best.

For $\beta > 4$ the SFF has additional ``spikes'' at higher integer values of $\tau$. This interesting phenomenon can be perhaps understood by thinking of the $\beta\to\infty$ limit. As $\beta$ grows larger the spectra become more regular: the level spacings are narrowly distributed around the average value of 1. In other words, the limit $\beta\to\infty$ is the zero temperature limit of the log gas, where spectra become exactly regular. For a completely regular spectrum $\lambda_n =\lambda_0 + n$, the SFF becomes
\begin{equation}
    \mathrm{SFF}(\tau) \to  \frac{1}{N}\vert\sum_{n=1}^N e^{2\pi i (\lambda_0+n) \tau}\vert^2 = \frac1N\left(\frac{\sin(N \pi \tau)}{\sin(\pi \tau)}\right)^2
\end{equation}
and therefore is expected to have narrow spikes of height $N$ and width $1/N$ at every integer $\tau$. In our plots what we're likely seeing is these spikes emerging in a continuous fashion for the SFF at large finite $\beta$. It is amusing to think of the log-singularity of the CSE SFF at $\tau = 1$ as being the onset of this large $\beta$ behavior.

Our final remark is that we did not investigate here the range $0\leq\beta\leq1$. The reason is that our Killip--Nenciu implementation starts suffering from numerical errors at small $\beta$ and has difficulty going below $\beta\approx0.3$.\footnote{The technical issue is that sampling the coefficients $\rho_k$ from the Beta distribution in eq.~\eqref{eq:KN_rho} begins giving values larger than 1 when $\beta$ is small due to round-off errors at the working precision, causing the algorithm to produce erroneous results.} The GBE and LBE implementations are more stable at small $\beta$ and can be used instead. That is beyond our current scope, but would be interesting to study, as the $\beta < 1$ regime interpolates between the GOE and Poisson-like statistics where the ramp of the SFF is not present.

\clearpage
\paragraph{Acknowledgments:} I thank Massimo Bianchi, Maurizio Firrotta, and Jacob Sonnenschein for their collaboration on the project from which this work has emerged and the many discussions along the way. This work was supported by an INFN postdoctoral fellowship and the INFN project ST\&FI “String Theory and Fundamental Interactions”.

\paragraph{Code and data availability:} The up-to-date package code is available on GitHub at \href{https://github.com/dorinw/beta_ensembles}{this URL}, with releases archived on Zenodo \cite{Weissman:2026be}. All results in sections \ref{sec:demo} and \ref{sec:applications} are reproducible from the provided code notebooks. The Zenodo release contains in addition the data files pertaining to the results cited in section \ref{sec:sff_exp}.

\appendix
\section{Addendum: Jacobi ensembles} \label{app:jacobi}
\begin{figure}[h!] \centering
\includegraphics[width=0.74\textwidth]{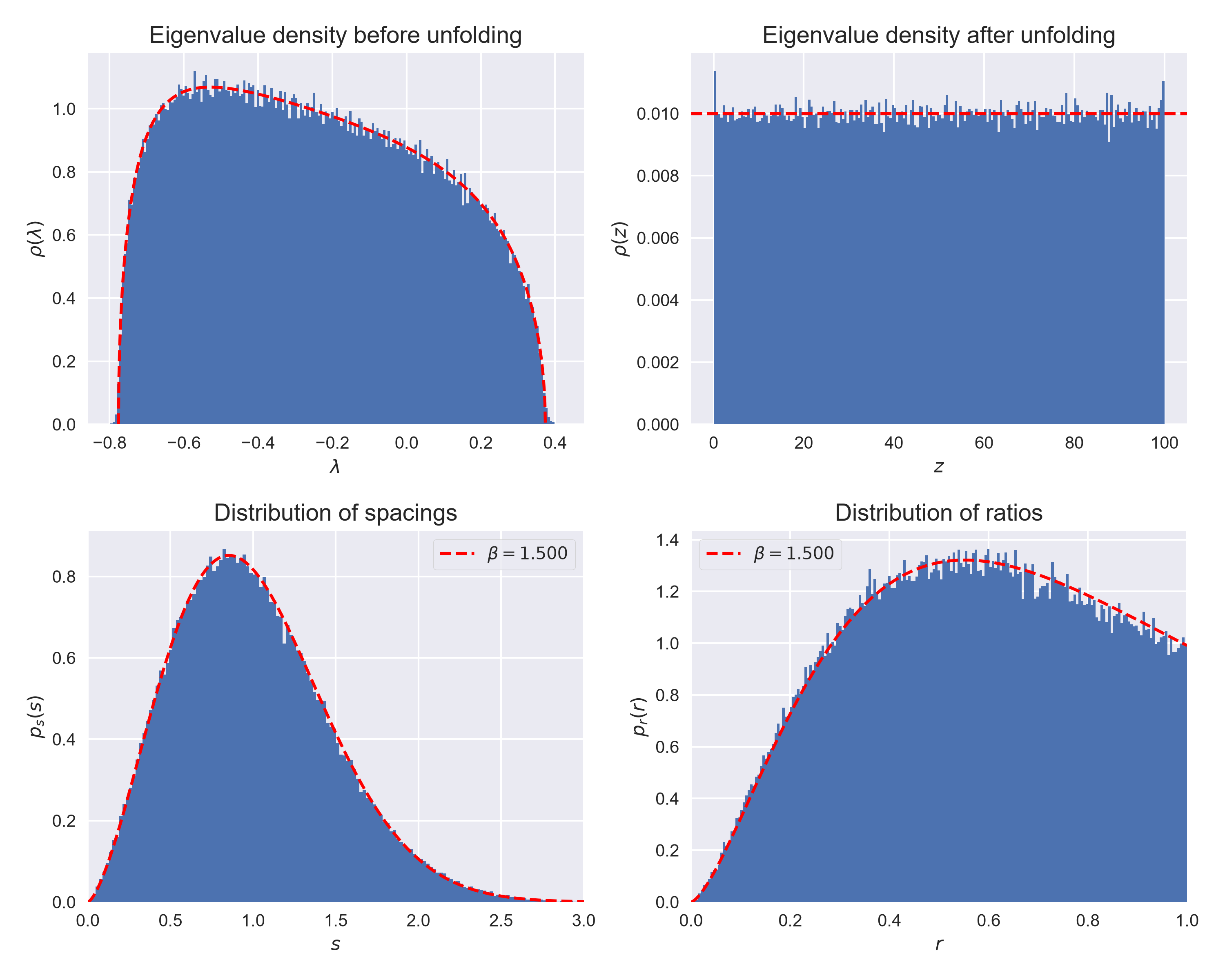}
    \caption{\label{fig:0_jbe_a_large} Results for the JBE with $\beta  = 1.5$, $a = 213$, $b=405$, $N = 100$, sample of 2000 random spectra.}
\end{figure}

\begin{figure}[ht!] \centering
\includegraphics[width=0.74\textwidth]{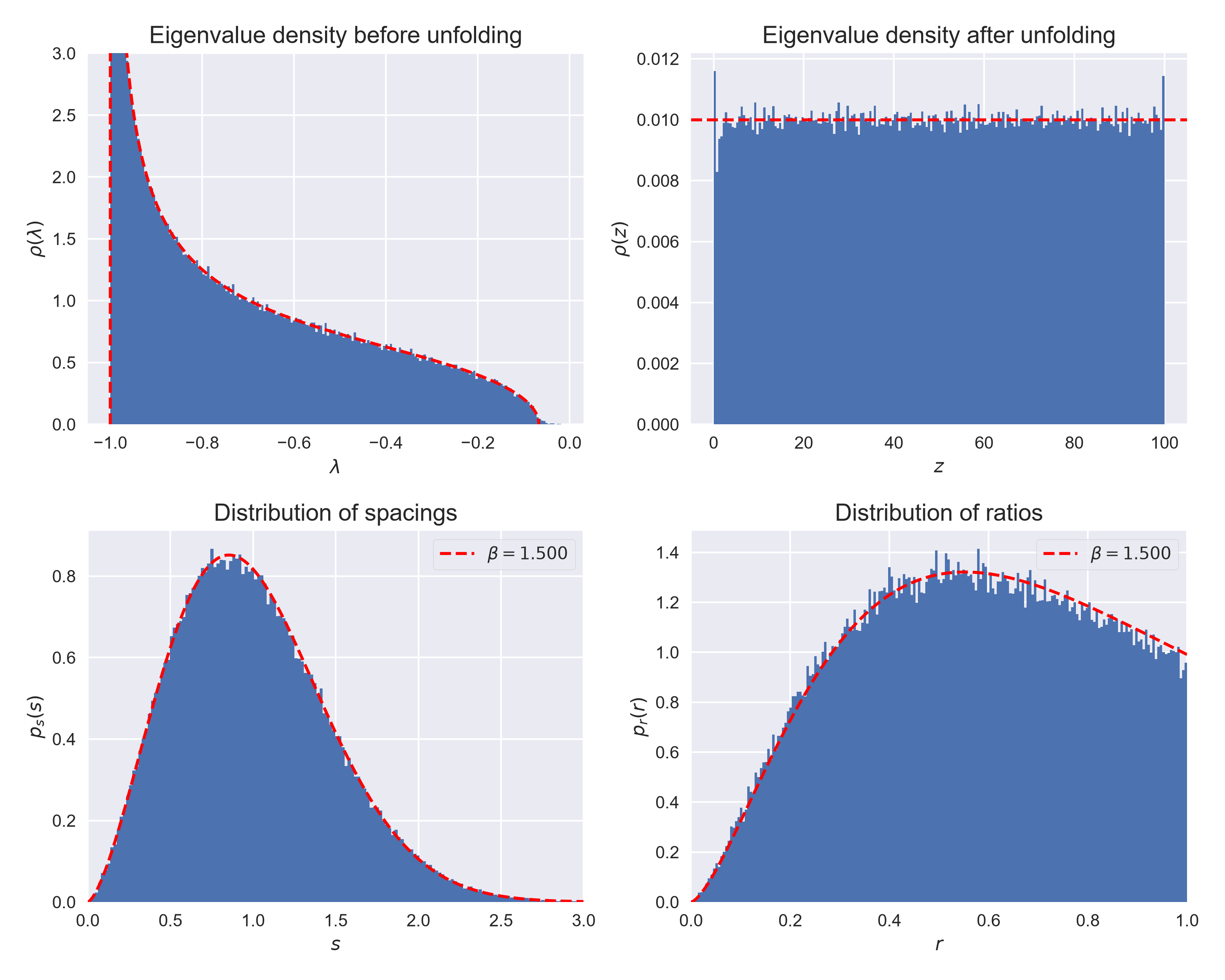}
    \caption{\label{fig:0_jbe_a_small} Results for the JBE with $\beta  = 1.5$, $a = -0.5$, $b=405$, $N = 100$, sample of 2000 random spectra.}
\end{figure}

The initial version of the package omitted the implementation of Jacobi ensembles.  To complete the demonstration of section~\ref{sec:demo}, we plot here the results for the Jacobi ensemble, on two samples of spectra. We take 2000 spectra with $N=100$ and set the parameters $\beta = 1.5$ and $b=405$. Between the two plots below we change the parameter $a$ from 213 in figure \ref{fig:0_jbe_a_large} to -0.5 in figure \ref{fig:0_jbe_a_small}. The spectral density changes accordingly (see eq.~ \eqref{eq:dist_rho_j}), but the distributions of spacings and spacing ratios match the familiar surmises for $\beta=1.5$. As always, the results can be reproduced from the notebook provided in the repository, and all parameters can be freely changed for further testing.

\bibliographystyle{JHEP}
\bibliography{SACS}

@article{Bianchi:2024fsi,
    author = "Bianchi, Massimo and Firrotta, Maurizio and Sonnenschein, Jacob and Weissman, Dorin",
    title = "{From spectral to scattering form factor}",
    eprint = "2403.00713",
    archivePrefix = "arXiv",
    primaryClass = "hep-th",
    reportNumber = "ITCP-IPP-2024/3",
    doi = "10.1007/JHEP06(2024)189",
    journal = "JHEP",
    volume = "06",
    pages = "189",
    year = "2024"
}

@ARTICLE{Killip:2004,
          doi = {10.48550/arXiv.math/0410034},
archivePrefix = {arXiv},
       eprint = {math/0410034},
 primaryClass = {math.SP},
title={Matrix models for circular ensembles},
  author={Killip, Rowan and Nenciu, Irina},
  journal={International Mathematics Research Notices},
  volume={2004},
  number={50},
  pages={2665--2701},
  year={2004},
  publisher={OUP}
}

@article{Cantero:2003,
  title={Five-diagonal matrices and zeros of orthogonal polynomials on the unit circle},
  author={Cantero, Maria J and Moral, Leandro and Vel{\'a}zquez, Luis},
  journal={Linear Algebra and its Applications},
  volume={362},
  pages={29--56},
  year={2003},
  publisher={Elsevier},
    eprint={math/0204300},
      archivePrefix={arXiv},
      primaryClass={math.CA}
}

@ARTICLE{Dumitriu:2002,
       author = {{Dumitriu}, Ioana and {Edelman}, Alan},
        title = "{Matrix models for beta ensembles}",
      journal = {Journal of Mathematical Physics},
         year = 2002,
        month = nov,
       volume = {43},
       number = {11},
        pages = {5830-5847},
          doi = {10.1063/1.1507823},
archivePrefix = {arXiv},
       eprint = {math-ph/0206043},
 primaryClass = {math-ph},
       adsurl = {https://ui.adsabs.harvard.edu/abs/2002JMP....43.5830D}
}

@article{Liu:2018hlr,
    author = "Liu, Junyu",
    title = "{Spectral form factors and late time quantum chaos}",
    eprint = "1806.05316",
    archivePrefix = "arXiv",
    primaryClass = "hep-th",
    reportNumber = "CALT-TH-2018-028",
    doi = "10.1103/PhysRevD.98.086026",
    journal = "Phys. Rev. D",
    volume = "98",
    number = "8",
    pages = "086026",
    year = "2018"
}

@book{Mehta:book,
  title={Random Matrices},
  author={Mehta, M.L.},
  isbn={9780080474113},
  edition={Third},
  url={https://www.elsevier.com/books/random-matrices/lal-mehta/978-0-12-088409-4},
  year={2004},
  publisher={Elsevier Science},
  doi={10.1016/S0079-8169(04)80091-6}
}

@article{Bianchi:2023uby,
    author = "Bianchi, Massimo and Firrotta, Maurizio and Sonnenschein, Jacob and Weissman, Dorin",
    title = "{Measuring chaos in string scattering processes}",
    eprint = "2303.17233",
    archivePrefix = "arXiv",
    primaryClass = "hep-th",
    doi = "10.1103/PhysRevD.108.066006",
    journal = "Phys. Rev. D",
    volume = "108",
    number = "6",
    pages = "066006",
    year = "2023"
}

@article{Atas:2013dis,
  title = {Distribution of the Ratio of Consecutive Level Spacings in Random Matrix Ensembles},
  author = {Atas, Y. Y. and Bogomolny, E. and Giraud, O. and Roux, G.},
  journal = {Phys. Rev. Lett.},
  volume = {110},
  issue = {8},
  pages = {084101},
  numpages = {5},
  year = {2013},
  month = {Feb},
  publisher = {American Physical Society},
  doi = {10.1103/PhysRevLett.110.084101},
  url = {https://link.aps.org/doi/10.1103/PhysRevLett.110.084101},
    eprint = "1212.5611",
    archivePrefix = "math-ph",
}

@book{Forrester:book,
url = {https://doi.org/10.1515/9781400835416},
title = {Log-Gases and Random Matrices (LMS-34)},
author = {Peter J. Forrester},
publisher = {Princeton University Press},
address = {Princeton},
doi = {doi:10.1515/9781400835416},
isbn = {9781400835416},
year = {2010},
lastchecked = {2023-01-30}
}

@article{Dyson1962,
  author  = {Dyson, Freeman J.},
  title   = {A Brownian-Motion Model for the Eigenvalues of a Random Matrix},
  journal = {Journal of Mathematical Physics},
  volume  = {3},
  number  = {6},
  pages   = {1191--1198},
  year    = {1962},
  doi     = {10.1063/1.1703862}
}

@article{Tekur:2018nme,
    author = "Tekur, S. Harshini and Bhosale, Udaysinh T. and Santhanam, M. S.",
    title = "{Higher-order spacing ratios in random matrix theory and complex quantum systems}",
    eprint = "1806.05958",
    archivePrefix = "arXiv",
    primaryClass = "quant-ph",
    doi = "10.1103/physrevb.98.104305",
    journal = "Phys. Rev. B",
    volume = "98",
    number = "10",
    pages = "104305",
    year = "2018"
}

@article{AbulMagd:1999,
  title={Wigner surmise for high-order level spacing distributions of chaotic systems},
  author={Abul-Magd, AY and Simbel, MH},
  journal={Physical Review E},
  volume={60},
  number={5},
  pages={5371},
  year={1999},
  publisher={APS},
  doi={10.1103/PhysRevE.60.5371}
}

@article{Weissman:2026be,
    title={{beta\_ensembles}: A Python package for continuous beta-ensembles},
  author={Dorin Weissman},
  journal={Zenodo},
  volume={},
  number={},
  pages={version 1.0},
  year={2026},
  publisher={Zenodo},
  doi={10.5281/zenodo.22127517}
}

@article{JMLR:v20:19-179,
  author  = {Guillaume Gautier and Guillermo Polito and R{{\'e}}mi Bardenet and Michal Valko},
  title   = {DPPy: DPP Sampling with Python},
  journal = {Journal of Machine Learning Research},
  year    = {2019},
  volume  = {20},
  number  = {180},
  pages   = {1--7},
  url     = {http://jmlr.org/papers/v20/19-179.html}
}

\end{document}